\documentclass[aps,pre,twocolumn,groupedaddress,amssymb,amsmath,showpacs]{revtex4-1}

\usepackage{amssymb,amsmath}
\usepackage{graphicx}
\usepackage{subfigure}
\usepackage{color}
\usepackage[hyperindex,breaklinks]{hyperref}

\newcommand{\rmd}{\mathrm{d}}
\newcommand{\nn}{\nonumber}
\newcommand{\du}{\dot{u}}

\newcommand{\tu}{\tilde{u}}

\newcommand{\hu}{\hat{u}}

\newcommand{\dw}{\dot{w}}
\newcommand{\ds}{\dot{s}}

\newcommand{\hl}{\hat{\lambda}}
\newcommand{\tl}{\tilde{\lambda}}

\newcommand{\tr}{\text{tr}}
\newcommand{\sgn}{\text{sgn}}
\newcommand{\tD}{\tilde{\Delta}}

\begin{document}

\title{Nonstationary dynamics of the Alessandro-Beatrice-Bertotti-Montorsi model}

\author{Alexander Dobrinevski}
\email[]{alexander.dobrinevski@lpt.ens.fr}
\author{Pierre Le Doussal}
%\email[]{ledou@lpt.ens.fr}
\author{Kay J\"org Wiese}
%\email[]{wiese@lpt.ens.fr}
\affiliation{CNRS-Laboratoire de Physique Th\'eorique de l'Ecole
Normale Sup\'erieure, 24 rue Lhomond, 75005 Paris
Cedex-France
\thanks{LPTENS is a Unit\'e Propre du C.N.R.S.
associ\'ee \`a l'Ecole Normale Sup\'erieure et \`a l'Universit\'e Paris Sud}
}
%\thanks{}

\date{\today\ %--- \jobname.tex version 
}

\begin{abstract}
We obtain an exact solution for the motion of a particle driven by a spring in a Brownian random-force landscape, the \textit{Alessandro-Beatrice-Bertotti-Montorsi} (ABBM) model. Many experiments on quasi-static driving of elastic interfaces  (Barkhausen noise in magnets, earthquake statistics, shear dynamics of granular matter) exhibit the  same universal behavior as this model. It also appears as a limit in the field theory of  elastic manifolds. %at or above the critical dimension 
Here we discuss  predictions of the ABBM model for monotonous, but otherwise arbitrary, time-dependent driving. Our main result is an explicit formula for the generating functional of particle velocities and positions. We apply this to derive the particle-velocity distribution following a quench in the driving velocity. We also obtain the joint avalanche size and duration distribution and the mean avalanche shape following a jump in the position of the confining spring. Such non-stationary driving is easy to realize in experiments, and provides a way to test 
 the ABBM model beyond the stationary, quasi-static regime. We study extensions to two elastically coupled layers, and to an elastic interface of internal dimension $d$, in the Brownian force landscape. The effective action of the field theory is equal to the action, up to 1-loop corrections obtained exactly from a functional determinant. This provides a connection to renormalization-group methods.
\end{abstract}


\maketitle

\section{Introduction}
The motion  of  domain walls in soft magnets   \cite{AlessandroBeatriceBertottiMontorsi1990,AlessandroBeatriceBertottiMontorsi1990b,Colaiori2008}, fluid contact lines on a rough surface \cite{LeDoussalWieseMoulinetRolley2009,MoulinetGuthmannRolley2002,RolleyGuthmannGombrowiczRepain1998}, or  strike-slip faults in geophysics  \cite{FisherDahmenRamanathanBenZion1997,BenZionRice1993,BenZionRice1997} can all be described on a mesoscopic level as motion of elastic interfaces driven through a disordered environment. Their response to external driving is not smooth, but exhibits discontinuous \textit{jumps} or \textit{avalanches}. Physically, these are seen e.g.~as pulses of Barkhausen noise in magnets \cite{Barkhausen1919,DurinZapperi2006b}, or  slip instabilities leading to earthquakes on geological faults \cite{Ruina1983,Dieterich1992,Scholz1998}. While the microscopic details of the dynamics are specific to each system, some large-scale features are  universal \cite{SethnaDahmenMyers2001}. The most prominent example are the exponents of the power-law distributions of avalanche sizes (for earthquakes, the well-known Gutenberg-Richter distribution  \cite{GutenbergRichter1944,GutenbergRichter1956,Kagan2002}) and durations.

The Alessandro-Beatrice-Bertotti-Montorsi (ABBM) model \cite{AlessandroBeatriceBertottiMontorsi1990} is a mean-field model for the dynamics of an interface in a disordered medium. It approximates a $d$-dimensional interface in a $d+1$-dimensional system, defined by a height function $u(x,t)$, by a single degree of freedom, its average height $u(t)=\frac{1}{L^d}\int \rmd^d x\,u(x,t)$. It satisfies the equation of motion
\begin{equation}
\label{eq:IntroABBM}
\partial_t u(t) = F\left(u(t)\right) - m^2\left[u(t)-w(t)\right].
\end{equation}
$w(t)$ is the external driving, and %, an example is driving with fixed velocity $w(t)=v t$.
$F(u)$  an {\em effective} random force, sum of the local pinning forces. In \cite{AlessandroBeatriceBertottiMontorsi1990}, 
it was postulated to be a Gaussian with the correlations of a Brownian motion 
\begin{equation}
\label{eq:CorrABBM}
\overline{\left[F(u_1)-F(u_2)\right]^2}=2\sigma |u_1-u_2|\ ,
\end{equation}
where $\sigma>0$ characterizes the disorder strength. 

This model has been analyzed in depth for the case of a constant driving velocity, i.e.~$w(t)=vt$ \cite{AlessandroBeatriceBertottiMontorsi1990,NarayanFisher1992,NarayanFisher1993,ZapperiCizeauDurinStanley1998,Colaiori2008,LeDoussalWiese2011,LeDoussalWiese2011c}. The distribution of avalanche sizes and durations was obtained by mapping \eqref{eq:IntroABBM} to a Fokker-Planck equation \cite{AlessandroBeatriceBertottiMontorsi1990,Colaiori2008}. The mean shape of an avalanche was also computed using this mapping \cite{PapanikolaouBohnSommerDurinZapperiSethna2011,ColaioriZapperiDurin2004,LeDoussalWiese2011}. These results agree well with numerous experiments on systems with long-range elastic interactions, realized e.g.~in certain classes of soft magnets, or in geological faults \cite{ZapperiCizeauDurinStanley1998,CizeauZapperiDurinStanley1997,DurinZapperi2000,Colaiori2008,FisherDahmenRamanathanBenZion1997}. 

However, long-range-correlated disorder as in \eqref{eq:CorrABBM} is a priori an  unphysical assumption for materials where the true microscopic disorder is, by
nature, short ranged. Hence in realistic systems, it can only arise as a model for the {\it effective disorder} felt by the interface. 
This guess, originally made by ABBM based on experiments, turns out to be very judicious.

In  \cite{ZapperiCizeauDurinStanley1998}, it was shown that the effective disorder for an interface with infinite-range elastic interactions is indeed given by \eqref{eq:CorrABBM}. This led to the wide belief that the ABBM model is a universal model for the center-of-mass of an interface in dimension $d$ at or above a certain upper critical dimension $d_c$ depending on the range of the elastic interactions in the system \footnote{Exactly at $d=d_c$, there are logarithmic corrections to the mean-field behaviour \cite{LeDoussalWiese2011,LeDoussalWiese2011c}, similar to those discussed in \cite{FedorenkoStepanow2003} and \cite{LeDoussalWiese2003} section VI.}.
Much of the popularity of the ABBM model is owed to this presumed universality. However only recently  this assumption was proven for short-ranged microscopic disorder using the Functional Renormalization Group (FRG) \cite{LeDoussalWiese2011,LeDoussalWiese2011c}, a method well suited to study interfaces (see \cite{WieseLeDoussal2006} for introduction and a short review). This proof required quasi-static driving $w(t)=vt$ with $v=0^+$.
Whether this property also holds for finite driving velocity $v>0$, and in that case up to which scale, requires further investigation. The same question for non-stationary driving also remains open.

There are some hints that non-stationary dynamics may require a different treatment. For example, avalanche size and duration exponents seem to vary over the hysteresis loop \cite{BertottiFiorilloSassi1981,DurinZapperi2006,SpasojevicBukvicMilosevicStanley1996}.

Related is the question of  static avalanches, i.e.~jumps in the order parameter
of the {\em ground state} upon variation of an external control parameter,
as e.g.~the magnetic field. This has been studied for elastic manifolds  via Functional RG methods \cite{LeDoussalWiese2008c,LeDoussalWiese2011b,LeDoussalMiddletonWiese2008}, and for spin glasses using Replica Symmetry Breaking \cite{LeDoussalMuellerWiese2010,LeDoussalMuellerWiese2011}.

In this paper, we discuss the results given by the ABBM model when the driving $w(t)$ is a monotonous, but otherwise arbitrary function of time. While this misses important and interesting physics of AC driving and the hysteresis loop \cite{Bertotti1998}, it is  much more general than the cases treated so far. We will give an analytic solution for arbitrary driving, and then specialize to examples such as the relaxation of the velocity $\du(t)$ after the driving is stopped, and the response to finite-size ``kicks'' in the driving force, $\dot w(t)=w_0 \delta(t)$. This should allow to clarify the range of the ABBM universality class by comparing these predictions to experiments and further theoretical work. Such non-stationary driving can easily be realized e.g.~in Barkhausen noise experiments, where $w(t)$ is the external magnetic field, and can be tuned as desired.

This paper is structured as follows. In section \ref{sec:SolABBM} we review the approach to the ABBM model through the Martin-Siggia-Rose (MSR) formalism. The MSR formalism maps disorder averages over solutions of the stochastic differential equation \eqref{eq:IntroABBM} to correlation functions in a field theory. In \cite{LeDoussalWiese2011,LeDoussalWiese2011c} this method was used to compute the Laplace transform of the $p$-point probability distribution of the velocity in the ABBM model, via the solution of a non linear ``instanton" equation. From it, the avalanche shape and duration distributions were  obtained for quasi-static driving, in agreement with the results of \cite{ColaioriZapperiDurin2004,PapanikolaouBohnSommerDurinZapperiSethna2011}. Here we extend the method of Ref. \cite{LeDoussalWiese2011,LeDoussalWiese2011c} and show that it 
is even more powerful: For any monotonous (but not necessarily stationary) driving $w(t)$ the resulting field theory can be solved exactly. We give an explicit formula for the generating functional of the particle velocity $\du$. In section \ref{sec:Examples} we apply this solution to several examples. In particular, we derive the  law for the decay of the velocity after the driving is stopped, which may easily be tested in experiments. In section \ref{sec:ABBMSpatial} we extend the method to variants of the ABBM model with additional spatial degrees of freedom. This includes the generalization of the ABBM model to a $d$-dimensional interface submitted to a quenched random force with the correlations of the Brownian motion, a model whose statics was studied in \cite{LeDoussalWiese2011b}. For this more general model, under monotonous driving, we show that the action of the field theory is not renormalized in any spatial dimension $d$. In section \ref{sec:FTPos}, we compute the generating functional for the particle position $u$, which is  more subtle than the one for the velocity $\du$. In sections \ref{sec:Generalizations} and \ref{sec:Conclusion}, we summarize the results and mention possible extensions. In particular, we explain why {\em  non-monotonous} motion requires a separate treatment, and does not  follow from the present results.

\section{Solution of the non-stationary ABBM model\label{sec:SolABBM}}
For understanding the physics of \eqref{eq:IntroABBM}, one would %primarily 
like to know the joint probability distribution for arbitrary sets of velocities $\du(t_1)...\du(t_n)$, averaged over all realizations of the random force $F$. This is encoded  in the generating functional
\begin{equation}
\label{eq:DefZ}
G[\lambda,w] = \overline{e^{\int_t \lambda(t)\du(t)}},
\end{equation}
where $\overline{\cdots}$ denotes disorder averaging. One then recovers e.g.~the generating function $\overline{e^{\lambda \du(t_0)}}$ of the distribution of $\du(t_0)$ by setting $\lambda(t)=\lambda \delta(t-t_0)$, and similarly for $n$-time correlation functions.
        
Our main result is an explicit formula for $G$ in the case of monotonous but non-stationary motion. Given the distribution of velocities $P_0(\du_i)$ at an initial time $t_i$, we claim that $G_{t_i}:=\overline{e^{\int_{t_i}^\infty \rmd t \lambda(t)\du(t)}}$ is 
        \begin{equation}
        \label{eq:SolZIC}
        G_{t_i}[\lambda,w]=e^{m^2\int_{t_i}^\infty \rmd t \tu(t) \dw(t)} \int_0^\infty \rmd \du_i P_0(\du_i)e^{\tu(t_i)\du_i}.
        \end{equation}    
Here $\tu(t)$ is the solution of an instanton equation  \cite{LeDoussalWiese2011,LeDoussalWiese2011c}:
        \begin{equation}
        \label{eq:EqUt}
        \partial_t \tu(t) - m^2 \tu(t) + \sigma \tu(t)^2 = -\lambda(t).
        \end{equation}
Boundary conditions are  $\tilde u(\infty)=0$; $\lambda(t)$ is assumed to vanish at infinity.     Note that $\tu(t)$ only depends on $\lambda(t)$, i.e.~the type of observable one is interested in, but not on the driving $w(t)$. The latter only enters in \eqref{eq:SolZIC}.

In the following, we are mostly interested in the case when the initial time $t_i\to -\infty$. Our observables will be local in time, so that $\lambda(t)$ decays quickly for $t\to \pm \infty$. Then, $\tu(t_i)\to 0$ and \eqref{eq:SolZIC} becomes independent of initial conditions, 
        \begin{equation}
        \label{eq:SolZ}
        G[\lambda,w]=e^{m^2\int_t\tu(t) \dw(t)}.
        \end{equation}
        To prove \eqref{eq:SolZ}, we first discuss how a closed equation for the velocity variable can be formulated. We then use the Martin-Siggia-Rose formalism to transform it to a field theory, and evaluate the resulting path integral to obtain \eqref{eq:SolZ}. Both steps use crucially the assumption of monotonous motion.
        
        \subsection{Velocity in the ABBM model}
The equation of motion for the velocity $\du(t)$ is obtained by differentiating \eqref{eq:IntroABBM}:
        \begin{equation}
                \label{eq:SDEVel}
                \partial_t \du(t) = \partial_t F\left(u(t)\right) - m^2\left[\du(t)-\dw(t)\right].
        \end{equation}
A priori, to determine the probability distribution of $\du(t)$, one needs  $\du(0)$ and $u(0)$, since the random force depends on the trajectory $u(t)$ and not just on $\du$. However, under the assumption that all trajectories are monotonous ($\du(t)\geq 0$ for all times $t$), the probability distribution of $\du(t)$ is independent of $u(0)$. Indeed, under this assumption, one can replace $\partial_t F(u(t))$ by a multiplicative Gaussian noise which only depends on $\du(t)$. More precisely, we can set $\partial_t F(u(t))=\sqrt{\du(t)}\xi(t)$ where $\overline{\xi(t)\xi(t')}=2\sigma\delta(t-t')$. To see this explicitly, consider the generating functional
        \begin{eqnarray}
        \nonumber
H[\lambda]&=&\overline{e^{\int_t \lambda(t) \partial_t F(u(t))}} \\
&=& e^{-\int_{t,t'} \lambda(t)\lambda(t')\frac{\sigma}{2}\partial_t\partial_{t'} |u(t)-u(t')|}.
\end{eqnarray}
Since $\du(t)\geq 0$ at all times, we know that \footnote{This is true even if
$\du(t)$ vanishes in some time interval, both sides being zero when both $t,t'$ belong to this interval,
or exhibits a jump.}
\begin{eqnarray}
\nonumber \partial_t\partial_{t'}|u(t)-u(t')|=\du(t)\partial_{t'}\text{sgn}\left(u(t)-u(t')\right)  \\
\label{eq:MonDelta}
= \du(t)\partial_{t'}\text{sgn}(t-t')=-2\du(t)\delta(t-t'),
\end{eqnarray}
and hence
\begin{equation}
H[\lambda] = e^{\int_{t} \lambda(t)^2\sigma\du(t)} = \overline{e^{\int_t \lambda(t) \sqrt{\du(t)}\xi(t)}}.
\end{equation}
Note that for monotonous driving, the monotonicity assumption $\du(t)\geq 0$ is enforced automatically if it at $t=t_0$ \footnote{This is a corollary of Middleton's theorem or ``no-passing rule'' \cite{Middleton1992,BaesensMacKay1998}. As a mathematical theorem, it is known since \cite{Hirsch1985}.}:
\begin{eqnarray}
\nonumber
& & \du(t_0)\geq 0, \dw(t)\geq 0 \; \text{for all}\;  t\geq t_0 \\
\label{eq:Middleton}
&\Rightarrow & \du(t)\geq 0 \; \text{for all}\; t\geq t_0.
\end{eqnarray}
In this way we see that for monotonous motion, \eqref{eq:SDEVel} is  a closed stochastic differential equation for the velocity $\du(t)$. Given an initial velocity distribution $P(\du(0))$, it can be solved without knowledge of the position $u(0)$.
        
\subsection{MSR field theory for the ABBM velocity\label{sec:SolMSR}}
The Martin-Siggia-Rose (MSR) approach allows us to express \eqref{eq:DefZ}, averaged over all realizations of $F$ in \eqref{eq:SDEVel} in a path integral formalism, following \cite{NarayanFisher1992,NarayanFisher1993,ChauveLeDoussalWiese2000a,LeDoussalWieseChauve2002,LeDoussalWiese2011,LeDoussalWiese2011c}.

Introducing the Wick-rotated MSR response field $\tu(t)$ and averaging over the disorder, one gets:
\begin{eqnarray}
\label{eq:MSRPathInt}
G[\lambda,w] &=& \int \mathcal{D}[\du,\tu] e^{-S[\du,\tu]+\int_t \lambda(t)\du(t)}, \\
\nonumber
S[\du,\tu] &=& \int_t \tu(t)\left[\partial_t \du(t) +m^2\left(\du(t)-\dw(t)\right) \right] \\
\nonumber
& & +\frac{\sigma}{2}\int_{t,t'}\partial_t \partial_{t'}|u(t)-u(t')|\tu(t)\tu(t').
\end{eqnarray}
Since we  consider only paths where $\du(t)\geq 0$ at all times, using \eqref{eq:MonDelta} we can rewrite the action as
\begin{eqnarray}
\nonumber
S[\du,\tu] = \int_t & & \left\{ \tu(t)\left[\partial_t \du(t) +m^2(\du(t)-\dw(t)) \right] \right.\\
\label{eq:MSRAction}
& & \left. - \sigma \du(t) \tu(t)^2  \right\}.
\end{eqnarray}
The key observation which allows to evaluate this exactly was first noted in \cite{LeDoussalWiese2011,LeDoussalWiese2011c}: The  action is linear in $\du(t)$. This means that the path integral over $\du$ can be evaluated, giving a $\delta$-functional. Instead of using this in the limit of $v\to 0$ as in \cite{LeDoussalWiese2011,LeDoussalWiese2011c}, one can write more generally:
\begin{eqnarray}
\nonumber 
G[\lambda,w] & = \int \mathcal{D}[\tu,\du] & e^{m^2 \int_t \tu(t)\dw(t)}\times \\ 
\nonumber & & \times
e^{\int_t \du(t)\left[\partial_t \tu(t) - m^2 \tu(t) + \sigma \tu(t)^2 + \lambda(t)\right]}  \\
%\label{eq:ZSolDelta}
\nonumber
&=\int \mathcal{D}[\tu] &e^{m^2 \int_t \tu(t)\dw(t)} \times \\
\nonumber
& \lefteqn{\times \delta\left[ \partial_t \tu(t) - m^2 \tu(t) + \sigma \tu(t)^2 + \lambda(t) \right],} &
\end{eqnarray}
This then reduces to \eqref{eq:SolZ} with $\tu(t)$ given by \eqref{eq:EqUt}. Note that the  Jacobian from evaluating the $\delta$-functional is independent of $w(t)$. We assume in the following that for $\dw(t)=0$ we have $\du=0$ and hence $G[\lambda,\dw=0]=\overline{e^{\int \lambda(t) \du(t)}}=1$ for any $\lambda$. Thus \eqref{eq:SolZ} is correctly normalized.

For the more rigorously minded reader, another derivation of \eqref{eq:SolZIC} and \eqref{eq:EqUt} is presented in appendix \ref{sec:AppDiscreteTime}. It avoids the use of path integrals with unclear convergence properties and takes into account the initial condition. 

\section{Examples\label{sec:Examples}}
\subsection{Stationary velocity distribution and propagator}
As a first application, let us re-derive the well-known probability distribution for the velocity in the case of stationary driving, $w(t)=v t$.

To obtain the generating function of the velocity distribution at $t_0$, we set $\lambda(t)=\lambda \delta(t-t_0)$ in \eqref{eq:DefZ}. The solution of \eqref{eq:EqUt} is \footnote{In the following we use dimensionless units: $m^2=1$, $\sigma=1$. Units can be restored by replacing 
$t \to t m^2$, $\tilde u \to  (\sigma/m^2) \tilde u$, $\lambda  \to  \lambda (\sigma/m^4)$
in the dimensionless solution for $\tu$. }
\begin{equation}
\label{eq:OneParticleUT}
\tu(t) = \frac{\lambda}{\lambda+(1-\lambda)e^{-(t-t_0)}}\theta(t < t_0).
\end{equation}
As already derived in \cite{LeDoussalWiese2011}, for $\dw(t)=v$ one gets
\begin{equation}
\int_t\tu(t) \dw(t) = -v \ln(1-\lambda),
\end{equation}
and hence $G(\lambda)=(1-\lambda)^{-v}$. This generating function yields the probability distribution
\begin{equation}
\label{eq:StatVelDist}
P(\du)= \frac{e^{-\du}\du^{-1+v}}{\Gamma(v)},
\end{equation}
which is the well-known result for the stationary velocity distribution \cite{AlessandroBeatriceBertottiMontorsi1990,Colaiori2008}.

Using the same method, we can obtain the 2-time velocity probability distribution. For $\lambda(t)=\lambda_1 \delta(t-t_1) +\lambda_2 \delta(t-t_2)$, with $t_1<t_2$, the solution of \eqref{eq:EqUt} is 
\begin{equation}
\label{eq:TwoParticleUt}
\tu(t) = \left\{ \begin{array}{lr}       
0 & t > t_2 \\       
\frac{1}{1 + \frac{1-\lambda_2}{\lambda_2}e^{t_2-t}} & t_1 < t < t_2 \\
\frac{1}{1 - \frac{\lambda_1\lambda_2 e^{t_1}-(1-\lambda_1)(1-\lambda_2)e^{t_2}}{(1+\lambda_1)\lambda_2 e^{t_1} +\lambda_1 (1-\lambda_2)e^{t_2}}e^{t_1-t}} & t < t_1.
\end{array}   \right.
\end{equation}

As already derived in \cite{LeDoussalWiese2011}, for $\dw(t)=v$ one gets
\begin{eqnarray}
\nonumber
\lefteqn{\int_t\tu(t) \dw(t) =} & & \\
\nonumber
& & -v \ln\left\{1-\lambda_1-\lambda_2 + \lambda_1\lambda_2\left[1-e^{-(t_2-t_1)}\right]\right\},
\end{eqnarray}
and using \eqref{eq:SolZ}
\begin{equation} 
\label{2timesZ}
G(\lambda_1,\lambda_2)=\left\{1-\lambda_1-\lambda_2 + \lambda_1\lambda_2\left[1-e^{-(t_2-t_1)}\right]\right\}^{-v}.
\end{equation}
Taking the inverse Laplace transform, we obtain the 2-time velocity distribution
\begin{equation}
\nonumber
P(\du_1,\du_2) = \frac{\sqrt{\du_1\du_2}^{-1+v}}{2\Gamma(v)\sinh \frac{\tau}{2} } I_{-1+v}\left(\frac{\sqrt{\du_1\du_2}}{\sinh \frac{\tau}{2}}\right)e^{\frac{v}{2}\tau -\frac{\du_1+\du_2}{1-e^{-\tau}}},
\end{equation}
where $\du_1:=\du(t_1)$, $\du_2:=\du(t_2)$, $\tau:=t_2-t_1 > 0$ and $I_\alpha$ is the modified Bessel function.
This formula generalizes the quasi-static result of \cite{LeDoussalWiese2011} to arbitrary $v$.
Dividing by the 1-point distribution $P(\du_1)$ given in \eqref{eq:StatVelDist}, one obtains a closed formula for the ABBM propagator for velocity $v>0$:
\begin{equation}
P(\du_2|\du_1)=\frac{\sqrt{\frac{\du_2}{\du_1}}^{-1+v}}{2\sinh \frac{\tau}{2} } I_{-1+v}\left(\frac{\sqrt{\du_1\du_2}}{\sinh \frac{\tau}{2}}\right)e^{\frac{v}{2}\tau -\frac{\du_1 e^{-\tau}+\du_2}{1-e^{-\tau}}}. 
\end{equation}
Using this result and the Markov property of equation \eqref{eq:SDEVel}, $n$-point correlation functions of the velocity can be expressed in closed form as products of Bessel functions.

\subsection{Velocity distribution after a quench in the driving speed \label{sec:VelDistQuench}}

Now let us consider a non-stationary situation. Assume that the domain wall is driven with a constant velocity $v_1 > 0$ for $t<0$,
which is changed to $v_2 \geq 0$ for $t>0$. One expects that the velocity distribution interpolates between the stationary distribution for $v_1$ at $t=0$ and the stationary distribution for $v_2$ for $t\rightarrow \infty$. In this section, we will compute its exact form for all times.

For the one-time velocity distribution, $\lambda(t)=\lambda \delta(t-t_0)$ and the solution of \eqref{eq:EqUt} is unchanged, given by \eqref{eq:OneParticleUT}.

\begin{figure}
        \centering
                \includegraphics[width=0.95\columnwidth]{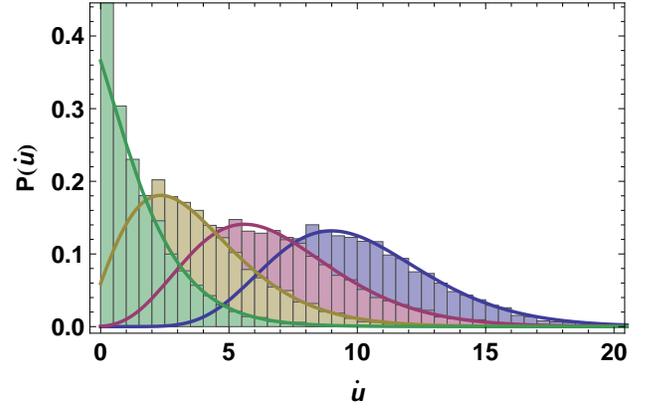}
        \caption{Decay of velocity distribution after driving was stopped. Curves (from right to left): Results for $t_0=0,0.4,1,2$ from \eqref{eq:ProbDistStopDriving}. Bar charts: Corresponding simulation results, averaged over $10^4$ trajectories. Initial driving velocity (for $t<0$) was $v=10$.}
        \label{fig:VelDistRelaxStopDriving}
\end{figure}

Now, using $\dw(t)=(v_1-v_2)\theta(-t)+v_2$, one gets
\begin{eqnarray}
\nonumber
\lefteqn{\int_t\tu(t) \dw(t)=} && \\
\nonumber 
&=& \int_{-\infty}^0 \frac{v_1\lambda\, dt}{\lambda+(1-\lambda)e^{-(t-t_0)}} 
+ \int_{0}^{t_0} \frac{v_2\lambda\, dt}{\lambda+(1-\lambda)e^{-(t-t_0)}} \\
\nonumber
&=& (v_1-v_2)\ln \left(1+\frac{\lambda}{1-\lambda}e^{-t_0}\right) - v_2 \ln (1-\lambda).
\end{eqnarray}
Thus, with the help of \eqref{eq:SolZ} 
\begin{equation}
\label{eq:ZQuench1}
G(\lambda)=\overline{e^{\lambda \du(t_0)}}=\left[1-\lambda(1-e^{-t_0})\right]^{v_1-v_2}\left(1-\lambda\right)^{-v_1}.
\end{equation}
Inverting the Laplace transform, one obtains
\begin{eqnarray}
\nonumber
P\left(\du(t_0)\right) &=& \frac{e^{-\du}\du^{-1+v_2}(1-e^{-t_0})^{v_1-v_2}}{\Gamma(v_2)} \times \\
\label{eq:ProbDistTwoVel}
& & \times \,_1 F_1\left(v_2-v_1,v_2,\frac{\du}{1-e^{t_0}}\right).
\end{eqnarray}
An interesting special case is when the driving is turned off at $t=0$, i.e.~$v_1=v$ and $v_2=0$. According to \eqref{eq:Middleton}, the particle will continue to move forward until it encounters the first zero of $\du=F(u)-m^2\left[u-w(0)\right]$. Correspondingly, we expect that the velocity distribution decays from the stationary probability distribution at $t\leq 0$ to a $\delta$-distribution at zero at $t\rightarrow \infty$. The explicit calculation for $\du := \du(t_0) $ yields:
\begin{eqnarray}
\label{eq:ProbDistStopDriving}
P(\du)&=&(1-e^{-t_0})^v \delta(\du) \\
\nonumber  &&+ e^{-\du-t_0 v}(e^{t_0}-1)^{-1+v} v \,_1 F_1\left(1-v,2,\frac{\du}{1-e^{t_0}}\right).
\end{eqnarray}
The $\delta(\du)$ term gives the probability that the motion has stopped at time $t_0$,
\begin{equation}
P(\du(t_0)=0)=(1-e^{-t_0})^v.
\end{equation}
As expected, this is zero at $t_0=0$ and tends to $1$ as $t_0\rightarrow \infty$.
Correspondingly, the distribution for the relaxation time $T$, i.e.~the time for the particle to stop moving from the stationary driving state at velocity $v$, is given by
\begin{equation}
\nonumber
P(T)=\frac{\partial}{\partial t_0}\bigg|_{t_0=T}P(\du(t_0)=0)=v e^{- T}(1-e^{- T})^{-1+v}.
\end{equation} 
The term in \eqref{eq:ProbDistStopDriving} not proportional to the $\delta$-function (once normalized) gives the conditional distribution of velocities assuming the particle is still moving. Its  form compares well to simulations, see figure \ref{fig:VelDistRelaxStopDriving}.

Using \eqref{eq:ZQuench1}, one also sees that the mean velocity interpolates exponentially 
between the old and the new value of the driving speed,
\begin{equation}
\label{eq:MeanVelDecay1}
\overline{\du_{t_0}}=\partial_\lambda \big|_{\lambda=0} G(\lambda) = v_2 + (v_1-v_2) e^{-t_0}.
\end{equation}
These results are valuable since they provide a tool to test the validity of the ABBM model in different experimental protocols. In application to Barkhausen noise, one could  perform experiments where the driving by the external magnetic field is stopped at some time. This would allow to verify e.g.~\eqref{eq:ProbDistStopDriving} experimentally, since the velocity  in our model is  the induced voltage  in a Barkhausen experiment. This would be one of the first checks on whether the good agreement between the ABBM theory and experiments persists in the non-stationary case. 

%{\red
\subsection{Non-stationary avalanches}

Using similar techniques, one can treat the case of a finite jump from $0$ to $w$ in the 
location of the confining harmonic well in \eqref{eq:IntroABBM}, $w(t)=w \theta(t)$ equivalent to a ``kick'' $\dw(t)=w\delta(t)$. For $t<0$ the particle is at rest, and the quench at $t=0$ triggers exactly one avalanche. Its size is given by $S=\int_0^\infty \du(t){\rm d} t$ and its duration $T$ by the first time when $\du(T)=0$. Note that this avalanche occurs as the non-stationary response to a kick of arbitrary size, a problem a priori different from the stationary avalanches studied previously \cite{Colaiori2008,LeDoussalWiese2011,LeDoussalWiese2011c} for small constant drive $\dw(t)=v =0^+$. In this section, we will derive the  distribution of avalanche sizes and durations for arbitrary kick sizes $w$.

\subsubsection{Preparation of the initial condition}\label{K1}
The assumption $G[\lambda,\dw=0]=1$ which we made in section \ref{sec:SolMSR} implies that the initial condition at  $t_i$, which is the lower limit of all time integrals in the action and in \eqref{eq:SolZ}, is \(\du (t_i)=0\). This means that the particle is exactly at rest for $t\geq t_i$ if $\dw(t)=0$ for $t\geq t_i$. Furthermore to assure that the particle will not revisit part of the trajectory, we demand\ \(u(t)\le u(t_i)\) for all \(t<t_i\). %\footnote{In the sense that %for any $t\geq t_i$, there are neither $t' \leq t_i$ nor $t' \geq t_i$ satisfying %$u_t=u_{t'}$.}. 
One protocol with which this can be enforced is: Start at some time $t_1 \ll t_i$ at an arbitrary position $u(t_1) \ll 0$, and take $w(t)=0$ for all $t\in [t_1,t_i]$. Then $\du(0)$ will be almost surely positive. Thus, between $t_1$ and $t_i$, the particle will move forward until it reaches the smallest $u$ where $F(u)-m^2 u = 0$. Since $t_1 \ll t_i$, almost surely it will reach this point before $t_i$ and thus be at rest at $t_i$.
This choice of initial condition is  equivalent to choosing a random configuration from the steady state for quasi-static driving at $v=0^+$.

\subsubsection{Duration distribution}

First, let us derive the exact distribution of  avalanche durations following a kick. The generating function for $P(\du(t_0))$ at time $t_0>0$ is obtained as in the previous section as
\begin{equation}
\label{eq:ZQuench2}
G(\lambda)=\overline{e^{\lambda \du(t_0)}}=\exp{\left(\frac{w\lambda}{\lambda+(1-\lambda)e^{t_0}}\right)}.
\end{equation}
Laplace inversion gives, denoting $\du := \du(t_0)$,
\begin{equation}
P(\du) =
\label{eq:ProbKick} 
 e^{- \frac{w+ e^{t_0} \du}{e^{t_0}-1}} \left[ \delta(\dot u) + \frac{1}{2 \sinh \frac{t_0}{2}} \sqrt{\frac{w}{\du}} 
I_1\left(\frac{\sqrt{\du w}}{\sinh \frac{t_0}{2}}\right) \right].  
\end{equation}
The mean velocity 
\begin{equation}
\overline{\du({t_0})}=\partial_\lambda \big|_{\lambda=0} G(\lambda) = w e^{-t_0}
\end{equation}
 decays in the same way  as in \eqref{eq:MeanVelDecay1} for  stopped driving. However
the  probability distributions of $\du(t_0)$ are different, as can be seen by comparing
\eqref{eq:ProbKick} and \eqref{eq:ProbDistStopDriving}. The probability that $\du(t_0)=0$, 
i.e.~that the avalanche has terminated at  time $T < t_0$, is  obtained by taking the
limit $\lambda\rightarrow -\infty$ in \eqref{eq:ZQuench2}, which gives the $\delta$-function
piece in \eqref{eq:ProbKick},\begin{equation}
\label{eq:DurationDist}
P(\du(t_0)=0)=P(T\leq t_0)=\exp{\left(-\frac{w}{e^{t_0}-1}\right)}.
\end{equation}
Note that this procedure requires \(P(\du<0)=0\) which is the case here. 

Correspondingly, the probability density for the avalanche duration $T$ is given by
\begin{equation} \label{pofT}
P(T) = \frac{\partial}{\partial t_0}\bigg|_{t_0=T} P(\du(t_0)=0) = \frac{w \exp{\left(-\frac{w}{e^T-1}\right)}}{\left(2\sinh\frac{T}{2}\right)^2}.
\end{equation}
We  observe that for infinitesimally small quenches $w$, one recovers -- up to a normalization factor -- the distribution obtained in \cite{LeDoussalWiese2011,LeDoussalWiese2011c} for avalanches at stationary, quasi-static driving, with the universal power law $T^{-2}$ for small times \cite{Colaiori2008}:
\begin{equation}
\label{rhoT}
 \rho(T) :=  \partial_w\big|_{w=0}P(T)= \frac{1}{\left(2\sinh\frac{T}{2}\right)^2}.
\end{equation}
Hence, the non-stationary character is not important in that limit. 

For finite $w>0$, the mean avalanche duration is obtained from \eqref{eq:DurationDist},
\begin{equation}
\nonumber
\overline{T}(w)=\gamma_E - e^w \text{Ei}(-w) + \log(w)\stackrel{w\rightarrow\infty}{\sim}\log w .
\end{equation}
It behaves as $\overline{T}(w) \sim w \ln(1/w)$ at small $w$ and diverges logarithmically for large $w$.
In the latter limit, the distribution of $\tilde T := T - \ln w$ approaches a  Gumbel distribution
\begin{equation} 
\nonumber
P(\tilde T) \approx  e^{- \tilde T} e^{- e^{- \tilde T}}
\end{equation}
on the interval $\tilde T \in [-\infty,\infty]$, as if the duration were given by the maximum
of $w$ independent random variables.

\subsubsection{Joint size and duration distribution}

One can now proceed to a more general case, and compute the joint distribution of avalanche durations and sizes. We  again calculate the generating function
\begin{equation}
\nonumber
G(\lambda_1,\lambda_2)=\overline{e^{\lambda_1 S + \lambda_2 \du(t_0)}}.
\end{equation}
where $S:= \int_0^\infty \du(t){\rmd t}$ is the avalanche size. 
The solution of \eqref{eq:EqUt} for $\lambda(t)=\lambda_1 + \lambda_2 \delta(t-t_0)$ is given by
\begin{eqnarray}
\nonumber
\tu(t) &=& \frac{1}{2}(1-\sqrt{1-4\lambda_1})+ \\
\nonumber
& & +\frac{e^{\sqrt{1-4\lambda_1}(t-t_0)}\sqrt{1-4\lambda_1}\lambda_2\theta(t_0-t)}{\sqrt{1-4\lambda_1}-\lambda_2[1-e^{\sqrt{1-4\lambda_1}(t-t_0)}]}.
\end{eqnarray}
Since the driving is $\dw(t)=w\delta(t)$, we obtain from \eqref{eq:SolZ}:
\begin{eqnarray}
G(\lambda_1,\lambda_2) &=& e^{w Z(\lambda_1,\lambda_2)}, \\
\nonumber
Z(\lambda_1,\lambda_2) &=&  \frac{1}{2}(1-\sqrt{1-4\lambda_1}) \\
&&  + \frac{e^{-\sqrt{1-4\lambda_1}t_0}\sqrt{1-4\lambda_1}\lambda_2}{\sqrt{1-4\lambda_1}-\lambda_2(1-e^{-\sqrt{1-4\lambda_1}t_0})}.
\qquad \end{eqnarray}
For $\lambda_2=0$ this gives the distribution of avalanche sizes $S$ for arbitrary kick size $w$, 
\begin{equation} \label{eq:PofSBM} 
P(S) = \frac{w}{2 \sqrt{\pi} S^{\frac{3}{2}}} \exp\left(- \frac{w^2}{4 S} - \frac{S}{4} + \frac{w}{2}\right).
\end{equation} 
As it should, this coincides with the distribution obtained for quasi-static driving, $v=0^+$ \footnote{Cf. formula (202) in \cite{LeDoussalWiese2009}. There, $S$ was  obtained as $S=u(w)-u(0)$ from the ``quasi-static'' position $u(w)$ where the velocity vanishes for the first time, i.e.~the largest solution of $u'(u)=F(u)-m^2 (u-w)=0$. Since $F(u)-m^2 (u-w)=0$ is a Brownian motion with drift, this is a standard first-passage problem. Thus, observing
the avalanche size $S$ alone, one cannot distinguish a non-stationary kick
from quasi-static driving.}.  

In the case of a non-stationary kick, we can obtain more information on the avalanche dynamics by considering the joint distribution of  avalanche sizes $S$ and durations $T$. 
As above, the probability that $\du(t_0)=0$ and hence the probability that the duration $T$ of the avalanche lies in the interval $]0;t_0]$ is given by the limit $\lambda_2\rightarrow -\infty$. Thus, the joint probability density $P(S,T)$ of sizes $S$ and durations $T$ satisfies 
\begin{eqnarray}
\nonumber && \int_0^\infty \rmd S\, \int_0^{t_0} \rmd T\, e^{\lambda_1 S} P(S,T) \\
\nonumber &&= \exp\left(\frac{w}{2}(1-\sqrt{1-4\lambda_1}) - w \frac{\sqrt{1-4\lambda_1}e^{-\sqrt{1-4\lambda_1}t_0}}{1-e^{-\sqrt{1-4\lambda_1}t_0}} \right).
\end{eqnarray}
Deriving with respect to $t_0$, we obtain 
\begin{eqnarray}
\nonumber
\lefteqn{\int_0^\infty \rmd S\, e^{\lambda S} P(S,T) =}& & \\
\label{eq:SOfTDist}
& & =\frac{w(1-4\lambda)e^{\frac{w}{2}\left(1-\sqrt{1-4\lambda}\coth\frac{T}{2}\sqrt{1-4\lambda}  \right)}}{\left(2\sinh \frac{T}{2}\sqrt{1-4\lambda}\right)^2},
\end{eqnarray}
which for $\lambda=0$ reproduces (\ref{pofT}). This implies the scaling form \footnote{
One notes the similarity of this formula with the one arising in the 
real space RG for the Brownian landscape in \cite{LeDoussalMonthusFisher1999}. It identifies with the square of Eq. (11) there, up to a global factor. It would be interesting to understand this connection further.}:
\begin{eqnarray} \label{st2}
&& P(S,T) = e^{- \frac{S}{4}} f(S/T^2)  \\
 && f(x) = {\rm LT}^{-1}_{s \to x} \frac{w e^{\frac{w}{2}} s e^{- \frac{w}{T^2} \sqrt{s} \coth \sqrt{s}}}{T^4 \left(\sinh \sqrt{s}\right)^2}
\end{eqnarray} 
Although no formula to invert the Laplace transform in a closed form is evident, one can, for example, calculate the mean avalanche size  for a fixed value of the avalanche duration,
\begin{eqnarray}
\nonumber
\overline{S}(T) &=& \frac{\int_0^\infty \rmd S\, S\,P(S,T)}{\int_0^\infty \rmd S\,P(S,T)} = \\
& = & \frac{4-w T - 4\cosh T +(2T+w)\sinh T}{\cosh T - 1}.\qquad 
\end{eqnarray}
As $w\rightarrow 0$, this has a well-defined limit
\begin{equation}
\label{eq:MeanSOfT}
 \overline{S}(T) = 2T \coth{\frac{T}{2}} - 4.
\end{equation}
Eq.~\eqref{eq:MeanSOfT} reproduces the expected scaling behaviour \cite{ZapperiCizeauDurinStanley1998,Colaiori2008}, $\overline{S}(T) \sim T^2$ for small avalanches. This is  apparent in (\ref{st2}), since the $e^{-\frac{S}{4}}$ factor can be neglected for small $S$. 
The new result in Eq.~\eqref{eq:MeanSOfT}   predicts   the deviations of large avalanches from this scaling, and shows that they obey $S\sim T$ instead. This  is in qualitative agreement with experimental observations on Barkhausen noise in polycristalline
FeSi materials \cite{DurinZapperi2006b,ColaioriZapperiDurin2004,Colaiori2008}. It would be interesting to test quantitative agreement of \eqref{eq:MeanSOfT} with experiments, as well.

We can also obtain the  large-$T$ behaviour at fixed $S$ (fixed $\lambda$) since in that limit\begin{eqnarray} 
&& \int_0^\infty \rmd S\, e^{\lambda  S} P(S,T)  \approx w (1- 4 \lambda) e^{w/2} e^{- \frac{1}{2} (w + 2 T) \sqrt{1 - 4 \lambda}}. \nn \\
&& \label{approx}
\end{eqnarray}
This implies
\begin{equation}
 P(S,T) \approx \frac{w (2 T+w) \left[(2 T+w)^2-6 S\right] e^{\frac{w}{2}-\frac{(2 T+w)^2}{4 S} - \frac{S}{4} } }{2 \sqrt{\pi } S^{7/2}}. 
  \label{approx2}
\end{equation} 
Note that (\ref{approx}) is  also valid at fixed $T$ and large negative $\lambda$, hence (\ref{approx2})
also gives the behaviour for $S\ll T^2$at fixed $T$. One notes some resemblance with (\ref{eq:PofSBM}). 

We  now consider  the limit of a small kick $w \to 0$. Eq.~\eqref{eq:SOfTDist} gives
\begin{eqnarray}
P(S,T) = w \rho(S,T) + O(w^2) ,
\end{eqnarray}
where $\rho(S,T)$ can be interpreted as an avalanche size and duration ``density", satisfying\begin{eqnarray}
\label{eq:ABBMSizeDurationDensity}
\int_0^\infty \rmd S\, e^{\lambda S} \rho(S,T) =\frac{(1-4\lambda)}{\left(2\sinh \frac{T}{2}\sqrt{1-4\lambda}\right)^2}.
\end{eqnarray} 
This Laplace transform can be inverted:
\begin{eqnarray}
&& \rho(S,T)  = e^{- S/4} \frac{1}{T^4} g(S/T^2) \label{a1} \\
&& g(x) = {\rm LT}_{s \to x}^{-1} \frac{s}{\left(\sinh \sqrt{s}\right)^2} = \frac{\rmd}{\rmd x} h(x) \label{a2} \\
&& h(x) =  \sum_{n=-\infty}^{+\infty} (1- 2 \pi^2 n^2 x) e^{- n^2 \pi^2 x} = \sum_{m=-\infty}^{\infty} \frac{2 m^2 e^{-\frac{m^2}{x}}}{\sqrt{\pi } x^{3/2}}. \nn
\end{eqnarray}
We have used $\sum_{n=-\infty}^{\infty} \frac{s- n^2 \pi^2}{(s+ n^2 \pi^2)^2} = 1/(\sinh \sqrt{s})^2$. Note that $\rho(S,T),$ as a size density, is normalized to 
$\int_0^\infty \rmd S \, \rho(S,T)= \rho(T)$, given in (\ref{rhoT}), since 
a fixed duration $T$ acts as  small avalanche-size cutoff. The 
total size density $\rho(S)=\int \rmd T\, \rho(S,T) = \frac{1}{2 \sqrt{\pi} S^{3/2}} \exp( - \frac{S}{4})$ 
is not normalized, since $w,$ which acts as a small-scale cutoff in (\ref{eq:PofSBM}), has been set to \(0\).

\begin{figure}
        \centering
                \includegraphics[width=0.9\columnwidth]{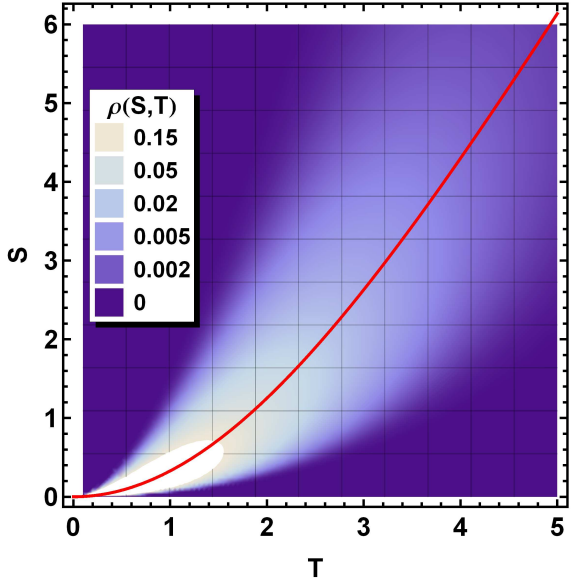}
        \caption{ Joint density $\rho(S,T)$ of avalanche sizes $S$ and durations $T$ in the ABBM model, obtained by numerical Laplace inversion of (\ref{eq:ABBMSizeDurationDensity},\ref{a1},\ref{a2}). The red line is the mean
        size $\bar S(T)$ for a fixed duration $T$ given in (\ref{eq:MeanSOfT}).}
        \label{fig:ABBMSizesDurations}
\end{figure}

Finally, note that \eqref{eq:SOfTDist} allows one to go further and compute any moment as well as, by numerical Laplace inversion, the full joint distribution $P(S,T)$. This is shown in figure \ref{fig:ABBMSizesDurations}.

 \subsubsection{Avalanche shape following a pulse}
We consider now the joint probability of velocities at two times $0< t_1<t_2$ following a pulse at time $t=0$.
By \eqref{eq:SolZ}, its generating function is
\begin{eqnarray}
\nn
\overline{e^{\lambda_1 \dot u(t_1) + \lambda_2 \dot u(t_2)}} = e^{w \tilde u(0)}, 
\end{eqnarray} 
where $\tilde u(0)$ is the 2-time solution (\ref{eq:TwoParticleUt}).
We are interested in $P(\dot u(t_1), \dot u(t_2)=0)$ obtained by taking $\lambda_2 \to -\infty$:
\begin{equation}
\nn 
 \int \rmd\dot u_1 e^{\lambda_1 \dot u_1} P(\dot u_1,0)=
\exp \left({ \frac{w}{1 - \frac{\lambda_1 e^{t_1}+(1-\lambda_1) e^{t_2}}{(1+\lambda_1) e^{t_1} -\lambda_1 e^{t_2}}e^{t_1}} }\right) 
\end{equation}
We use that $\mbox{LT}^{-1}_{s \to u} e^{d + \frac{a}{b+s}} = e^d [ \sqrt{\frac{a}{u}} I_1(2 \sqrt{a u}) e^{-b u} + \delta(u)]$
with $d=-\frac{w}{e^{t_1}-1}$, $a=we^{t_1}/(e^{t_1}-1)^2$ and $b=\frac{1}{e^{t_2-t_1}-1} + \frac{1}{1- e^{-t_1}}$. 
Taking $\partial_{t_2}$ and setting $t_2=T$ we find the joint probability distribution of the avalanche duration $T$
and the velocity $\dot u(t_1)=\dot u_1$,
\begin{eqnarray}
\nn
 P(\dot u_1,T) &=& - \partial_{t_2} b e^d  \sqrt{a \dot u_1} I_1(2 \sqrt{a \dot u_1}) e^{-b \dot u_1}  |_{t_2=T} \\
\nn
& =& \frac{1}{\left[2 \sinh(\frac{T-t_1}{2})\right]^2} \frac{\sqrt{w \dot u_1}}{2 \sinh\frac{t_1}{2}} 
I_1\left(\frac{\sqrt{w \dot u_1}}{\sinh \frac{t_1}{2}}\right) \\
&& \times e^{-\frac{w}{e^{t_1}-1} - \left(\frac{1}{e^{T-t_1}-1} + \frac{1}{1- e^{-t_1}}\right) \dot u_1}.
\end{eqnarray}
Dividing by $P(T)$ given in (\ref{pofT}), we find the conditional probability for the velocity distribution at
$t_1$ for an avalanche of duration $T$. In particular, we get the average avalanche shape,
\begin{equation}
\overline{\dot u(t_1)}_T = \frac{ 4 \sinh(\frac{t_1}{2}) \sinh(\frac{T-t_1}{2})}{\sinh(\frac{T}{2})} 
+ w \left[\frac{ \sinh(\frac{T-t_1}{2}) }{ \sinh(\frac{T}{2})} \right]^2
\end{equation}
For $w \to 0$ one recovers the stationary avalanche shape obtained in \cite{PapanikolaouBohnSommerDurinZapperiSethna2011,LeDoussalWiese2011}. On the other hand, avalanches following a pulse of size $w > 0$ have an asymmetric shape, since
$\dot u(t=0^+)=w$. This should provide an elegant way to discriminate between the
two situations experimentally. 

\subsection{Power spectral density and distribution of Fourier modes} % and oscillatory driving}
In signal analysis, an important observable used to characterize a time series is the \textit{power spectral density} $P(\omega)$ defined as
\begin{equation}
\label{eq:PSDDef}
P(\omega) := \lim_{T\to\infty}\frac{1}{T}\overline{\left|\int_{-T/2}^{T/2} e^{i\omega t}\left[\du(t) - \overline{\du(t)}\right] \mathrm{d}t \right|^2}.
\end{equation}
This gives a measure for the abundance of the frequency component $\omega$ in the time series $\du(t)$. For a stationary signal where the 2-time velocity correlation function only depends on the time difference, \eqref{eq:PSDDef} is  equal to its Fourier transform:
\begin{equation}
\label{eq:PowerSpFT}
P(\omega) = \int_{-\infty}^{\infty}  e^{i\omega t}\, \overline{\du(0) \du(t)}^c\,\mathrm{d}t.
\end{equation}
For driving with constant velocity $w(t)=vt$, one knows \cite{Colaiori2008,LeDoussalWiese2009} $\overline{\du(0) \du(t)}^c = v e^{-|t|}$ and hence the power spectrum for the velocity in the ABBM model is
\begin{equation}
\label{eq:PowerDensity}
P(\omega) = \frac{2v}{1+\omega^2}.
\end{equation}
We can now proceed further and obtain the probability density of each Fourier component. We consider \eqref{eq:SolZ} with 
$\lambda(t)=\lambda\cos\omega t\, \theta(T-t) \theta(t)$ where $T$ is a large-time cutoff. To solve \eqref{eq:EqUt} with this choice of $\lambda$, we substitute $\tu(t)=\frac{1}{2}+\frac{\phi'(t)}{\phi(t)}$ giving \textit{Mathieu's equation},
\begin{equation}
\nonumber
\phi''(t) - \left(\frac{1}{4}-\lambda \cos \omega t\right)\phi(t)=0.
\end{equation}
This is to be solved with the boundary condition $\tilde u(T)=0$, i.e. \(\phi'(T)=-\frac12 \phi(T)\).

The general solution is a linear combination of two \textit{Floquet} solutions
\begin{equation}
\label{eq:SolMathieu}
\phi(t) = e^{\mu t}P_1(t) + e^{-\mu t}P_2(t),
\end{equation}
where $P_{1,2}(t)$ are periodic functions.
%One observes that for $0 < t \ll T$, the solution tends to a Floquet solution $\phi(t) = e^{i\mu t}P(t)$, where $P(t)$ is a periodic function of time. 
$\mu=\mu(\lambda, \omega)$ is related to the conventionally defined \textit{Mathieu characteristic exponent} $\nu(a,q)$ (in the notation of \cite{AbramowitzStegun1965}) by 
\begin{equation}
\nonumber
\mu = \frac{\omega}{2i}\,\nu\!\left(-\frac{1}{\omega^2},\frac{2\lambda}{\omega^2}\right).
\end{equation}
%MathieuCharacteristicExponent
When $\lambda$ is real and close to $0$, $\mu$ is real, has the same sign as $\lambda$, and is odd in \(\lambda\). Thus, for $0<t \ll T$, the solution $\phi(t)$ given in \eqref{eq:SolMathieu} is dominated by the exponentially decaying term
\begin{equation}
\nn
\phi(t) \approx  e^{- \mu(|\lambda|,\omega) t} P(t),
\end{equation}
with $P(t)=P_{1,2}(t)$, depending on the sign of $\lambda$. Thus, for $0 < t \ll T$ we have
\begin{equation}
\tu(t) = \frac{1}{2} -\mu(|\lambda|,\omega) + \frac{P'(t)}{P(t)}.
\end{equation}
In order to evaluate  \eqref{eq:SolZ}, one needs to integrate \(\tilde u(t)\)
over \(t\) from 0 to \(T\). Since $P(t)$ is periodic, its contribution vanishes for each  period
\begin{equation}
\int_s^{s+\frac{2\pi}{\omega}} \tu(t)\,\rmd t = \frac{2\pi}{\omega}\left[\frac{1}{2} - \mu(|\lambda|,\omega)\right], \quad 0 < s \ll T.
\end{equation}
For constant driving, $w(t)=vt$ and $T \gg \frac{2\pi}{\omega}$, one thus obtains using \eqref{eq:SolZ}
\begin{equation}
\label{eq:FourierDensity}
%\overline{\frac{1}{T}e^{\lambda\int_0^T \du(t)\cos\omega t}} = e^{v\frac{1}{T}\int_0^T \tu(t)} = e^{v\left[\frac{1}{2}-i \frac{\omega}{2}\nu\left(-\frac{1}{\omega^2},\frac{2\lambda}{\omega^2}\right) \right]}.
\overline{e^{\lambda\int_0^T \du(t)\cos\omega t}} = e^{v\int_0^T \tu(t)} = e^{vT\left[\frac{1}{2} + i \frac{\omega}{2} \nu\left(-\frac{1}{\omega^2},\frac{2|\lambda|}{\omega^2}\right) \right] + o(T)}.
\end{equation}
As expected by symmetry, this is an even function in $\lambda$. It remains real as long as the Mathieu exponent $\nu$ is purely imaginary, which is the case for $|\lambda|<\lambda_c (\omega)$. One can interpret the corresponding Mathieu functions as Schr\"{o}dinger wavefunctions in the periodic potential
\begin{equation}
\nn
V(x) = \frac{1}{4} - \lambda \cos(\omega x)
\end{equation}
The region $|\lambda|<\lambda_c (\omega)$ is the region where the energy $E=0$ is outside the energy band(s) of this potential, and all wave-functions are evanescent. At $\lambda=\pm \lambda_c$, one has $\nu=0$ and for $|\lambda|>\lambda_c$, i.e.~outside the ``band gap'', the expectation value on the left-hand side of \eqref{eq:FourierDensity} does not exist. This indicates that the distribution of $\int_0^T \du(t)\cos\omega t$ has exponential tails for any $\omega>0$. The exponent of this tail can be computed in terms of the so-called Mathieu characteristic values $a_r$ and $b_r$ \cite{AbramowitzStegun1965}. 

Furthermore, from \eqref{eq:FourierDensity} one observes the scaling behaviour of the cumulants
\begin{equation}
\overline{\left(\int_0^T \du(t)\cos\omega t \right)^n}^c \sim T,
\end{equation} 
which reminds of the central limit theorem.

Taking two derivatives of \eqref{eq:FourierDensity} with respect to $\lambda$, and using $\partial_q^2\big|_{q=0^+} \nu(-b^2,q)=\frac{-i}{2b(b^2+1)}$, one verifies once more \eqref{eq:PowerDensity}. However, \eqref{eq:FourierDensity} goes beyond that and gives the full probability distribution of each frequency component of the time series $\du(t)$.

With this, we conclude our examples on the ``classical'' ABBM model and move to generalizations which can be treated by our method, as well.

\section{ABBM model with spatial degrees of freedom \label{sec:ABBMSpatial}}

An interesting generalization of the ABBM model \eqref{eq:IntroABBM} is a model with spatial degrees of freedom, (e.g.~an extended elastic interface in  dimension $d>0$), but subject to the same kind of disorder as in the ABBM model, i.e.~a pinning force correlated as a random walk.

An interface was studied in \cite{LeDoussalWiese2011} for quasi-static driving and it was found that the global motion (i.e.~the motion of the center-of-mass of the interface) is unchanged by the elastic interaction. An instanton equation for the other Fourier modes was derived, but solving it remained a challenge. 

Here we extend these results to arbitrary driving velocity. We first study the simpler case of only two 
elastically coupled particles, and present a direct argument to show that the center of mass is not affected by the elastic interaction and is the same as for a single particle, i.e.~model \eqref{eq:IntroABBM} in a rescaled disorder. For two particles the instanton equation is simpler and more amenable to analytic studies, which  allows us to see how local properties (such as the velocity distribution of a single particle) are modified. In the last
part we come back to the interface and show a non-renormalization property of the theory valid for any driving 
velocity.

\subsection{Two elastically coupled particles in an ABBM-like pinning-force field}
The model we  analyze in this section is a 2-particle version of \eqref{eq:IntroABBM}:
\begin{eqnarray}
\nonumber 
\partial_t u_1(t) &=& F_1(u_1(t)) - m^2\left[u_1(t)-w(t)\right] \\
\nonumber
& & + k\left[u_2(t)-u_1(t)\right], \\
\nonumber
\partial_t u_2(t) &=& F_2(u_2(t)) - m^2\left[u_2(t)-w(t)\right] \\
\label{eq:TwoPartABBMDef}
& & + k\left[u_1(t)-u_2(t)\right].
\end{eqnarray}
We assume $F_1(u_1)$, $F_2(u_2)$ to be independent Gaussian processes with correlations as in \eqref{eq:CorrABBM}, i.e.
\begin{equation}
\nonumber
\overline{\left[F_1(u)-F_1(u')\right]^2}=\overline{\left[F_2(u)-F_2(u')\right]^2}=2\sigma |u-u'|.
\end{equation} 

\subsubsection{Center-of-mass motion}
From \eqref{eq:TwoPartABBMDef}, we obtain the  equation of motion for the center-of-mass velocity $\ds(t)=\frac{1}{2}[\du_1(t)+\du_2(t)]$:
\begin{equation}
\label{eq:TwoParticleCenterOfMassEOM}
\partial_t \ds(t) = \frac{1}{2}\partial_t\left[F_1(u_1(t))+F_2(u_2(t))\right] - m^2\left[\ds(t)-\dw(t)\right].
\end{equation}
To better understand the effective noise term $\partial_t\left[F_1(u_1(t))+F_2(u_2(t))\right]$, let us compute its generating functional,
\begin{eqnarray}
\nonumber
G[\lambda]&=&\overline{e^{\int_t \lambda(t) \frac{1}{2}\partial_t\left[F_1(u_1)+F_2(u_2)\right]}} \\
\nonumber
&=& e^{-\int_{t,t'} \lambda(t)\lambda(t')\frac{\sigma}{8}\partial_t\partial_{t'} (|u_1(t)-u_1(t')|+|u_2(t)-u_2(t')|)}.
\end{eqnarray}
Using monotonicity \cite{Middleton1992,BaesensMacKay1998} of the trajectories \eqref{eq:MonDelta}, we obtain
\begin{equation}
\nonumber
G[\lambda] = e^{\int_{t} \lambda(t)^2\frac{\sigma}{4}\left[\du_1(t)+\du_2(t)\right]} = e^{\int_{t} \lambda(t)^2 \frac{\sigma}{2}\ds(t)}.
\end{equation}
Note that this is the same generating function as for a  random pinning force $F(s(t))$ with correlations
\begin{equation}
\overline{\left[F(s)-F(s')\right]^2}=\sigma|s-s'|.
\end{equation}
Thus, we can re-write \eqref{eq:TwoParticleCenterOfMassEOM} as 
\begin{equation}
\partial_t \ds(t) = \partial_t F(s) - m^2\left[\ds(t)-\dw(t)\right],
\end{equation}
with a rescaled disorder amplitude $\sigma'=\frac{\sigma}{2}$, reducing it to the same form as \eqref{eq:SDEVel}. 

This argument extends straightforwardly to any number of elastically coupled particles, and to the continuum limit. Thus, we observe that the dynamics of the center of mass of an extended interface in a pinning-force field, which is correlated as a random walk, is \textit{equivalent} to the 1-particle ABBM model \eqref{eq:IntroABBM}.

\subsubsection{Single-particle velocity distribution}
\begin{figure}
        \centering
                \includegraphics[width=0.95\columnwidth]{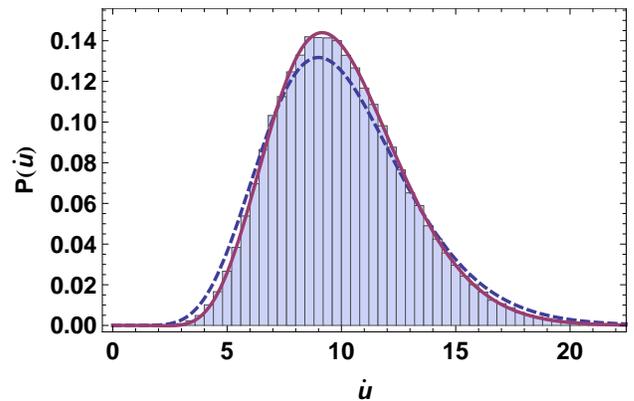}
        \caption{Single-particle velocity distribution $P(\du)$ in the 2-particle toy model for weak elasticity. Histogram: Numerical results from simulations for $k=0.2$. Dashed line: Stationary distribution in absence of elastic coupling ($k=0$). Solid (red) line: $\mathcal{O}(k)$ result from \eqref{eq:TwoParticleSingleVelDist}.}
        \label{fig:DiffDistSmallK}
\end{figure}

On the other hand, observables that can not be described solely in terms of the center of mass are more complicated. In order to obtain the joint distribution of the particle velocities $\du_1(t), \du_2(t)$ one may follow the same route as in section \ref{sec:SolMSR}. We start from  
\begin{eqnarray}
\nonumber
G[\lambda_1,\lambda_2,w]&=&\overline{e^{\int_t\lambda_1(t)\du_1(t)+\lambda_2(t)\du_2(t)}} \\
&=& e^{m^2\int_t \left[\tu_1(t)+\tu_2(t)\right]\dw(t)},
\end{eqnarray}
where $\tu_1$, $\tu_2$ are solutions of the coupled nonlinear differential equations
\begin{eqnarray}
\nonumber
-\partial_t \tu_1(t) + m^2\tu_1(t)+ k\left[\tu_1(t)-\tu_2(t)\right] & &\\
\nonumber  - \sigma\tu_1(t)^2 & =&  \lambda_1(t), \\
\nonumber
-\partial_t \tu_2(t) + m^2\tu_2(t) + k\left[\tu_2(t)-\tu_1(t)\right]& & \\
\nonumber  - \sigma\tu_2(t)^2 &=& \lambda_2(t).
\end{eqnarray}
In contrast to \eqref{eq:EqUt}, these can not be solved in a closed form even for simple choices of $\lambda_{1,2}$. However, one can obtain a perturbative solution for small $k$ around $k=0$.
%\begin{widetext}
To give a simple example, one obtains for monotonous driving $w(t)=v t$ and one-time velocity measurements $\lambda_{1,2}(t)=\lambda_{1,2} \delta(t)$:
\begin{align}
\label{eq:TwoParticleZl1l2}
&\lefteqn{G(\lambda_1,\lambda_2) = \left[(1-\lambda_1)(1-\lambda_2)\right]^{-v}}\nn\\
&\times \bigg\{1
+vk(\lambda_1-\lambda_2)\left[-\frac{\ln(1-\lambda_1)}{\lambda_1(1-\lambda_2)} +\frac{\ln(1-\lambda_2)}{\lambda_2(1-\lambda_1)}\right]\nn\\
&\qquad \ \, + \mathcal{O}(k^2) \bigg\}.
\end{align}
%\end{widetext}
where we use    rescaled units where $k$ denotes $k/m^2$ in the original units. As one expects from the previous section, the correction of order $k$ vanishes if one considers the center-of-mass motion, $\lambda_1=\lambda_2$. If, on the other hand, one considers the 1-particle velocity distribution, i.e.~takes $\lambda_2=0$, one gets
\begin{equation}
G(\lambda_1,0)=(1 - \lambda_1)^{-v(1 + k)}\left[1-vk\frac{\lambda_1}{1 - \lambda_1} +\mathcal{O}(k^2)\right].
\end{equation}
The Laplace transform can be inverted, giving
\begin{eqnarray}
\label{eq:TwoParticleSingleVelDist}
P(\du_1)  &= & \frac{e^{-\du_1}\du_1^{-1+v}}{\Gamma(v)} \times \\ 
\nonumber
 & & \times
\left\{1+ k\left[v-\du_1+ v\ln{\du_1}-v\psi(v)\right] + \mathcal{O}(k^2)\right\},
\end{eqnarray}
where $\psi(x)=\frac{\Gamma'(x)}{\Gamma(x)}$ is the digamma function. Simulations for small $k$ confirm this result (see figure \ref{fig:DiffDistSmallK}). The next order in $k$ can likewise be calculated, however the resulting expressions are complicated and not very enlightening.

A non-trivial consequence of \eqref{eq:TwoParticleSingleVelDist} is that the power-law exponent of the distribution $P(\du_1)$ for small velocities changes from $\du^{-1+v}$ to $\du^{-1+v(1+k)}$.

\subsection{Continuum limit and non-renormalization property\label{sec:NonRen}}
Let us now consider a $d$-dimensional interface in a $d+1$-dimensional medium with a generic elastic kernel $g_{xy}$, such that in Fourier $g^{-1}_{q=0}= m^2$. Local elasticity corresponds to $g_q^{-1}=q^2 + m^2$. The corresponding generalization of \eqref{eq:IntroABBM} is
\begin{equation}
\label{eq:EOMSpatial}
\partial_t u_{xt} = F(u_{xt},x) - \int_y g^{-1}_{xy}u_{yt} + \tilde \lambda_{xt}.
\end{equation}
For the remainder of this section, we write function arguments as subscripts in order to simplify notations (i.e.~$u_{xt}:=u(x,t)$).
The source $\tl_{xt} \geq 0$ for the field $\tu$ is a positive driving, and is related to the velocity of the center of the quadratic well $\dw$ by $\tl_{xt}=g^{-1}_{xx'} \dw_{x't}$.

The pinning force is chosen gaussian and uncorrelated in $x$,
\begin{eqnarray} \label{defDE} 
\overline{F(u,x) F(u',x')} = \delta^{d}(x-x') \Delta(u,u').
\end{eqnarray} 
In the $u$ direction, analogously to (\ref{eq:CorrABBM}), we assume Brownian correlations, i.e.~uncorrelated increments:
$\partial_u \partial_{u'} \Delta(u,u') = \delta(u-u')$. This does not fix $F$ uniquely, with e.g.~two possible explicit choices in
(\ref{eq:DeltaBM}) and (\ref{eq:DeltaStat}) below. However, differences only arise for the   position $u$ but not for the velocity $\dot u$, as will be discussed below.

Let us  write the MSR partition sum in presence of sources,
\begin{eqnarray}
\nonumber
&& G[\lambda, \tilde \lambda] = \int \mathcal{D} [\du] \mathcal{D} [\tu] e^{- S[\du ,\tu ] + \int_{xt} \lambda_{xt} \du_{xt} + \int_{xt} \tl_{xt} \tu_{xt} }.
\end{eqnarray}
The generalization of the MSR action \eqref{eq:MSRAction} to this situation is\begin{eqnarray}
\nonumber
S[\du ,\tu ]  &=& \int_{xt} \tu_{xt}\left(\partial_t \du_{xt} + \int_y g^{-1}_{xy} \du_{yt} - \sigma\du_{xt}\tu_{xt} \right).\\
\label{eq:MSRActionSpatial}
\end{eqnarray}
To arrive at (\ref{eq:MSRActionSpatial}) we have again assumed forward-only trajectories $\dot u_{x t} \geq 0$, guaranteed if $\tilde \lambda_{xt} \geq 0$ and $\dot u_{x t_i} \geq 0$ at some
large negative initial time $t_i$.

The solution in section \ref{sec:SolMSR} generalizes straightforwardly to 
\begin{equation}
\label{eq:SolZSpatial}
G[\lambda,w] = \overline{e^{\int_{xt}\lambda_{xt}\du_{xt}}} = e^{\int_{xt}\tu^{(s)}_{xt}[\lambda]\tilde \lambda_{xt}},
\end{equation}
where $\tu^{(s)}[\lambda]$ is defined as the solution of
\begin{equation}
\label{eq:ABBMSGDeftus}
\partial_t \tu^{(s)}_{xt} - \int_y g^{-1}_{xy}\tu^{(s)}_{yt} + \sigma(\tu^{(s)}_{xt})^2 = -\lambda_{xt}.
\end{equation}
In principle, this can be used to compute any observable of the $d$-dimensional theory. In practice, the equation \eqref{eq:ABBMSGDeftus} for $\tu$ is hard to solve analytically for most cases.

In the remainder of this section, instead of discussing specific examples, we show a conceptual consequence of \eqref{eq:SolZSpatial}: \textit{The action \eqref{eq:MSRActionSpatial} does not renormalize}. The effective action $\Gamma$ is equal to the microscopic action $S$ in any dimension $d$. 

According to \eqref{eq:SolZSpatial}, the generating functional for connected graphs $W[\lambda,\tl]$ evaluates to
\begin{equation}
\nonumber
W[\lambda,\tl] = \ln{ G[\lambda,\tl] } = \int_{xt} \tu_{xt}^{(s)}[\lambda] \tl_{xt}.
\end{equation}
To perform the Legendre transform from $W$ to the effective action $\Gamma$ \cite{LeDoussalWiese2006a}, we introduce new fields $\du_{xt}[\lambda, \tilde \lambda]$, and $\tu_{xt}[\lambda, \tilde \lambda]$, defined by
\begin{eqnarray}
\label{eq:ABBMSGDeftu}
\tu_{xt} &=& \frac{\delta W[\lambda,\tl]}{\delta \tl_{xt}} = \tu_{xt}^{(s)}[\lambda], \\
\label{eq:ABBMSGDefu}
\du_{xt} &=& \frac{\delta W[\lambda,\tl]}{\delta \lambda_{xt}} = \int_{x't'} \frac{\delta\tu_{x't'}^{(s)}[\lambda]}{\delta \lambda_{xt}} \tl_{x't'}. \qquad
\end{eqnarray}
Here and below we drop the functional dependence on the sources when no ambiguity arises. Eq.\
\eqref{eq:ABBMSGDeftu} shows that $\tu_{xt}^{(s)}[\lambda]$ is really the field $\tu_{xt}$ appearing in the effective action, hence \eqref{eq:ABBMSGDeftus} allows to express the field $\lambda_{xt}$ (on which $W$ depends) in terms of $\tu_{xt}$ (on which $\Gamma$ depends).

We can now write down the effective action $\Gamma[u,\tu]$:
\begin{eqnarray}
\nonumber
\Gamma[\du,\tu] &=& \int_{xt} \du_{xt} \lambda_{xt} + \int_{xt} \tu_{xt} \tl_{xt} - W \\
\nonumber
&=& \int_{xt} \du_{xt} \lambda_{xt} \quad \text{since} \quad W= \int_{xt} \tu_{xt} \tl_{xt} \\
\nonumber
&=& -\int_{xt} \du_{xt} \left(\partial_t \tu_{xt} - \int_y g^{-1}_{xy}\tu_{yt} + \sigma\tu_{xt}^2\right)\\
\nonumber
&=& \int_{xt} \tu_{xt} \left(\partial_t \du_{xt} + \int_y g^{-1}_{xy}\du_{yt} - \sigma\du_{xt}\tu_{xt}\right) \\
&=& S[\du ,\tu ] .
\end{eqnarray}
This is exactly the same as  the bare action $S$ in \eqref{eq:MSRActionSpatial}. This non-renormalization of the action for the particle velocity in ABBM-like disorder is also consistent with a 1-loop calculation using functional RG methods (see appendix \ref{sec:AppFDet}). It is a very non-trivial statement, and shows that, {\it in some sense}, the MSR field theory for monotonous motion in ABBM-like disorder is exactly solvable in any dimension. The monotonicity assumption implies that the derivatives arising in the formulae above must be performed in the neighborhood of a strictly positive driving source $\tl_{xt} >0$. Using the relationship 
\begin{equation}
\label{newu2}
\du_{xt}[\lambda, \tilde \lambda] = \frac{\overline{\du_{xt} \exp\left[ \int_{x't'} \lambda_{x't'} \du_{x't'}\right]}}{\overline{\exp\left[ \int_{x't'} \lambda_{x't'} \du_{x't'}\right]}},  
\end{equation}
(where the average is performed in presence of $\dot{w}=\tilde \lambda$)
one sees that \eqref{eq:ABBMSGDefu} maps positive $\tl$ onto positive $\du$. 
%Hence the velocity $\du$ appearing in $\Gamma$ is likewise positive. 
On the other hand, the condition $\tl \geq 0$ can be expressed using $\tl_{xt}=\frac{\delta \Gamma}{\delta \tu_{xt}}$ as
\begin{equation}
\label{eq:CondPosLT}
\tu_{xt} \leq \frac{\partial_t \du_{xt} + \int_y g^{-1}_{xy}\du_{yt}}{2\sigma \du_{xt}}\ .
\end{equation}
We conclude that the effective action $\Gamma[\du,\tu]$ is given by the bare action $S$ in the sector of the theory where $\du \geq 0$ and \eqref{eq:CondPosLT} holds as a necessary condition. In no way this implies that $\Gamma=S$ for values of the fields where this monotonicity assumption does not hold. The case of non-monotonous motion and/or non-monotonous driving is highly non-trivial and will be studied elsewhere. 

In the following section, we shall see how this result generalizes to the field theory of the position $u(t)$, where the relationship between $S$ and $\Gamma$ is slightly more complicated.

\section{Field theory for the position variable\label{sec:FTPos}}
So far, we have  considered observables that can be expressed in terms of the ABBM velocity $\du(t)$, or in case of a manifold \(\du(x,t)\). Here we consider the position   \(u(x,t)  \) itself. One can then formulate the MSR path integral in terms of $u$ and \( \hat u,\) analogous to \eqref{eq:MSRPathInt}. This was done for a $d$-dimensional interface in short-ranged disorder in  \cite{LeDoussalWiese2011c}, as a starting point for a $d_c-d$-expansion. Here we focus on the simpler and solvable case of the ABBM model, where the MSR path integral reads\begin{eqnarray}
\nonumber
G[\lambda,w]&=&\overline{e^{\int_t \lambda(t) u(t)}} = \int \mathcal{D}[u,\hu] e^{-S[u,\hu]+\int_t \lambda(t)u(t)}, \\
\nonumber
S[u,\hu] &=& \int_t \hu(t)\left[\partial_t u(t) +m^2(u(t)-w(t)) \right] \\
\label{eq:MSRPathIntPos}
& & - \frac{1}{2}\int_{t,t'}\Delta\left(u(t),u(t')\right)\hu(t)\hu(t').
\end{eqnarray}
Here, $\Delta(u,u') = \overline{F(u)F(u')}$ is the disorder correlation function. One mathematically simple choice is to assume the random force $F(u)$ to be a one-sided Brownian motion and restrict to $u>0$:
\begin{equation}
\label{eq:DeltaBM}
\Delta(u,u')=2\sigma \min (u,u') = \sigma(u + u' - |u-u'|).
\end{equation}
Another common choice is the two-sided version, i.e.~a Brownian motion on the full real $u$ axis
pinned at $F(u=0)=0$. With either choice, however, the random force is non-stationary and one loses
statistical translation invariance. This is unnatural for certain applications, for example approximating extended elastic interfaces above the critical dimension. In this context, one chooses a stationary variant of \eqref{eq:DeltaBM},
\begin{equation}
\label{eq:DeltaStat}
\Delta(u,u') = \Delta(u-u')= \Delta(0)-\sigma|u-u'|.
\end{equation}
Since a stochastic process $F(u)$ can only satisfy \eqref{eq:DeltaStat} for all $u$ in some limit, we
always assume \eqref{eq:DeltaStat} to be regularized at large $|u-u'|$.

For observables that can be expressed in terms of the velocity $\du$, only $\partial_t \partial_{t'} \Delta(u(t),u(t'))$ enters the MSR action (cf. section \ref{sec:SolMSR}). Hence, choosing \eqref{eq:DeltaBM} or \eqref{eq:DeltaStat} yields the same result \eqref{eq:MSRAction}. However, the choice does matter if one is interested in observables depending on the position, like the mean pinning force $f_p:=m^2\overline{[u(t)-w(t)]}$.

In contrast to the velocity theory discussed in previous sections, fixing a distribution of positions $u(t_i)$ as the initial condition is problematic. Indeed, in general one cannot exclude that this initial condition leads to backward motion $\dot u(t_i) <0$ for
some realizations of the disorder. Hence for the stationary Brownian landscape \eqref{eq:DeltaStat} we will choose $t_i = - \infty$ and assume that the driving $\dot w(t) \geq 0$ is such that at fixed times the initial condition is forgotten, as  discussed in section \ref{K1}. We claim that then
\begin{eqnarray}
\label{eq:SolZPosStat}
 G[\lambda,w] &=& \overline{e^{\int_t \lambda(t) u(t)}} \\
& =& 
e^{m^2 \int_t  \hu(t) w(t) +  \frac{\Delta(0)}{2m^4}\left[\int_t  \lambda(t)\right]^2}\left[1-\frac{\sigma}{m^4} \int_t  \lambda(t)\right], \nn
\end{eqnarray}
where all time integrals are over $]-\infty,\infty[.$ The function $\hu(t)=-\partial_t \tu(t)$ where $\tu(t)$ is  solution of
\begin{equation}
        \label{eq:EqUtPosStat}
        \partial_t \tu(t) - m^2 \tu(t) + \sigma \tu(t)^2 - \sigma \tu(t)\tu(-\infty)= - \int_{t'>t}\lambda(t').
\end{equation}
In the particular case of the one-sided Brownian landscape \eqref{eq:DeltaBM} we  only consider the initial condition $u(t_i)=0$. Since $F(0)=0$ in that case, for $w(t_i) \geq 0$ and $\dot w(t) \geq 0$ the motion will be forward. Then
the generating function $G[\lambda,w]$ in \eqref{eq:MSRPathIntPos} takes a form  analogous to \eqref{eq:SolZ}
\begin{equation} 
\label{nice} 
G[\lambda,w]=\overline{e^{\int_{t>t_i} \lambda(t) u(t)}}=e^{m^2\int_{t>t_i} \hu(t) w(t)},
\end{equation}
where $\hu(t)=-\partial_t \tu(t)$ and $\tu(t)$ is  solution of \eqref{eq:EqUt}. In the remainder of this section, we shall prove the above statements 
%\eqref{eq:SolZPosStat} 
and then apply these formulae to  determine the distribution of the single-time particle position $u(t)$.

\subsection{Generating functional for stationary Brownian potential\label{sec:PosFTStat}}

Using the assumption of monotonous motion, the disorder term in the action \eqref{eq:MSRPathIntPos} can be rewritten as
\begin{eqnarray}
\nonumber
\lefteqn{\frac{1}{2} \int_{t,t'}\Delta(u(t),u(t'))\hu(t)\hu(t') = }& & \\
\nonumber
& & =\frac{\Delta(0)}{2}\left[\int_t \hu(t)\right]^2-\sigma \int_{t,t'} u(t) \hu(t) \hu(t') \sgn(t-t').
\end{eqnarray}
Following the same approach as in section \ref{sec:SolMSR}, evaluating the path integral over $u(t)$ in \eqref{eq:MSRPathIntPos} yields 
%\begin{widetext}
\begin{align}\label{eq:ZPos1}
& \int \mathcal{D}[\hu] e^{m^2 \int_t \hu(t)w(t) + \frac{\Delta(0)}{2}\left[\int_t \hu(t)\right]^2}\nn\\
&\times \delta\left( \partial_t \hu(t) - m^2 \hu(t) + \sigma \hu(t)\! \int_{t'} \hu(t') \text{sgn}(t'{-}t) + \lambda(t) \right).
\end{align}
%\end{widetext}
Thus  
\begin{equation}
        \label{eq:SolZPos}
        G[\lambda,w]=\mathcal{N}e^{m^2\int_t \hu(t) w(t)  + \frac{\Delta(0)}{2}\left[\int_t \hu(t)\right]^2},
\end{equation}
where $\hu(t)$ is  solution to the equation
%\begin{equation}
        %\label{eq:EqUh1}
        %\partial_t \hu(t) - m^2 \hu(t) + 2\sigma \hu(t) \int_{t'>t} \hu(t') = -\lambda(t)
%\end{equation}
%For the stationary case (correlator given by \eqref{eq:DeltaStat}), \eqref{eq:EqUh1} is changed to 
\begin{equation}
        \label{eq:EqUhStat}
        \partial_t \hu(t) - m^2 \hu(t) + \sigma \hu(t) \int_{t'} \hu(t') \text{sgn}(t'-t) = -\lambda(t).
\end{equation}
Substituting $\tu(t):=\int_t^\infty \hu(t){\rm  d}t$, one recovers \eqref{eq:EqUtPosStat}. $\tu(-\infty)$ is obtained from 
\begin{equation}
-m^2 \int_{-\infty}^\infty \hu(t) = -m^2 \tu(-\infty) = -\int_{-\infty}^\infty \lambda(t')\, \rmd t'.
\end{equation}
Note that  $\tu(-\infty)$ vanishes for $\lambda$ such that $\int_{t'} \lambda(t') =0$. These are exactly those observables which can be expressed in terms of the velocity (or, equivalently, position differences). 

As in section \ref{sec:SolABBM}, $\mathcal{N}$ in \eqref{eq:SolZPos} is the normalization of the path integral and the Jacobian of the operator inside the $\delta$-functional in \eqref{eq:ZPos1}. It is independent of $w(t)$, but we cannot fix its value at $w(t)=\rm const$ as we did for the velocity theory in section \ref{sec:SolABBM}: Even if one keeps $w=\rm const$ for a long time, the distribution of $u$ will remain nontrivial (unlike the distribution of $\du$, which will become $\delta(\du)$). Here, to fix $\mathcal{N}$ we compare to the disorder-free solution ($\sigma=0$) for which the trajectory $u(t)$ is deterministic and satisfies \eqref{eq:SolZPos} with $\mathcal{N}=1$. Hence, we can write $\mathcal{N}$ as a ratio of functional determinants arising from the $\delta$-functional,
\begin{equation} \label{eq:DefJacobian}
\mathcal{N}^{-1} =\frac{\det(\partial_t-m^2 - \Sigma^T)}{\det(\partial_t-m^2)} = \det (1+R\Sigma).
\end{equation}
Here, $R$ is the disorder-free propagator
\begin{equation}
\label{eq:PropFree}
R:=\left(\partial_t+m^2\right)^{-1} \Rightarrow R_{t_1,t_2}=\theta(t_1-t_2)e^{-m^2(t_1-t_2)},
\end{equation}
and $\Sigma$ is the disorder ``interaction'' term, or "self-energy"
\begin{eqnarray}
\nonumber
 \Sigma^T_{t_2,t_1} = \Sigma_{t_1,t_2} &=&  \sigma \delta(t_1-t_2) \int_{t'} \hu(t') \text{sgn}(t_1-t') \\
\label{eq:DefSigma}
& & +\sigma \hu(t_2) \text{sgn}(t_2-t_1).
\end{eqnarray}
By explicit computation (see appendix \ref{sec:AppFDet}), one verifies that 
\begin{equation}
\nonumber
\tr \left(R\Sigma\right)^n = -\left[- \frac{\sigma}{m^2}\int_t \hu(t)\right]^n,
\end{equation}
and hence 
\begin{equation}
\nonumber
\det (1+R\Sigma) = \exp \tr \ln(1+R\Sigma)= \left(1-\frac{\sigma}{m^2}\int_t \hu(t)\right )^{\!\!-1}.
\end{equation}
From \eqref{eq:EqUhStat}, one further knows that $\int_t \hu(t) = \frac{1}{m^2}\int_t \lambda(t)$.
 
In total, this proves the expression \eqref{eq:SolZPosStat} for the stationary case, 
\begin{equation}
G[\lambda,w]= e^{m^2\int_t \hu(t) w(t) +  \frac{\Delta(0)}{2m^4}\left[\int_t \lambda(t)\right]^2}\left[1-\frac{\sigma}{m^4}\int_t \lambda(t)\right]
\end{equation}
One sees again that for observables expressed in terms of the velocity, where $\int_t \lambda(t)=0$, the simpler expression \eqref{eq:SolZ} is recovered.

In the language of perturbative field theory, the non-trivial functional determinant signifies non-vanishing 1-loop diagrams \footnote{However, two- and higher-loop corrections still vanish.}. This is in contrast to the theory for the velocity (section \ref{sec:SolMSR}), where all observables were given by tree-level diagrams. These loop corrections mean that the non-renormalization property discussed in section \ref{sec:NonRen} has to be amended when considering the particle position in a stationary potential. After renaming the driving $w$ to $\hl=m^2 w$, the source for the field $\hu$, the generating functional for connected correlation functions becomes
\begin{equation}
\nonumber
W[\lambda, \hl] = \int_t \hu_t[\lambda] \hl_t + \frac{\Delta(0)}{2 m^4}\left(\int_t \lambda_t\right)^2 
+ \ln\left(1{-} \frac{\sigma}{m^4} \int_t \lambda_t\right),
\end{equation}
%Taking a functional derivative w.r. to $\hl$, one sees that $\hu_t[\lambda]$ is still the field $\hu$ appearing in the effective action $\Gamma$. 
where $\hu_t[\lambda]$ is  solution of (\ref{eq:EqUhStat}).
Following the same procedure as in section \ref{sec:NonRen}, one obtains the effective action 
\begin{eqnarray}
\nonumber
\lefteqn{\Gamma[u,\hu]=} & & \\
\nonumber & =& \int_t u_t  \left[-\partial_t \hu(t) + m^2 \hu(t) - \sigma \hu(t) \int_{t'} \hu(t') \text{sgn}(t'-t)\right] \\
\nonumber
& &- \frac{\Delta(0)}{2}\left[\int_t \hu(t)\right]^2- \ln\left[1- \frac{\sigma}{m^2} \int_t \hu(t)\right] \\
&=& S[u,\hu] - \ln\left[1- \frac{\sigma}{m^2} \int_t \hu(t)\right].
\end{eqnarray}
We thus see that the property $\Gamma=S$ seen for the velocity theory is only changed by a simple contribution from the 1-loop corrections. The equal-time part of the $\hu^n$ term of these loop corrections coincides with a previous result in \cite{FedorenkoLeDoussalWiese2006}. 

In fact, this calculation can be extended to the $d$-dimensional interface with elastic kernel $g_q$ of section \ref{sec:NonRen}. There too it ensures that for the position theory, and monotonous driving, $\Gamma$ differs from $S$ only via the logarithm of a (one-loop) functional determinant. Thus, 2- and higher-loop corrections to correlation functions and the effective action vanish. Its expression is particularly simple in the case of a uniform $\lambda_{xt}=\lambda(t)$ leading to a uniform saddle point $\hat u_{xt}=\hat u(t)$:
\begin{eqnarray}
&& W_{\text{1-loop}} = L^d \int \frac{\rmd^d q}{(2 \pi)^d} \ln\left[ 1 - \frac{\sigma g_q}{m^2} \int_t \lambda(t) \right] \\
&& \Gamma - S|_{{\rm uniform }\,\hat u} = L^d \int \frac{\rmd^d q}{(2 \pi)^d} \ln\left[ 1 - \sigma g_q \int_t \hat u(t) \right]  \nn
\end{eqnarray}
 $L^d$ is the volume of the system. Details and a more general discussion are given in appendix \ref{sec:AppFDet}, appendix \ref{frg} and \cite{LeDoussalWiese2011c}.

\subsection{One-sided Brownian potential\label{sec:PosFTBrownian}}

It is instructive to give for comparison the solution for the simpler case of the correlator \eqref{eq:DeltaBM}. Using the assumption of monotonous motion, the disorder term in the action \eqref{eq:MSRPathIntPos} can be rewritten as 
\begin{equation}
\sigma\int_{t,t'}\min \left(u(t),u(t')\right)\hu(t)\hu(t') = 2\sigma \int_t u(t) \hu(t) \int_{t'>t} \hu(t')
\end{equation}
Following the same approach as in section \ref{sec:SolMSR}, evaluating the path integral over $u(t)$ with initial
condition $u(t_i)=0$ in \eqref{eq:MSRPathIntPos} yields
equation (\ref{nice}), where $\hu(t)$ is  solution to the equation
\begin{equation}
        \label{eq:EqUh1}
        \partial_t \hu(t) - m^2 \hu(t) + 2\sigma \hu(t) \int_{t'>t} \hu(t') = -\lambda(t).
\end{equation}
Note that as in section \ref{sec:SolMSR}, the initial condition $u(t_i)=0$ ensures that $G[\lambda,w=0]=1$. 
Hence the functional determinant analogous to \eqref{eq:DefJacobian} is equal to $1$ in this case. This is
also checked by a direct calculation in Appendix \ref{sec:AppFDet}. For $\lambda(t)$ non-vanishing only around $t \gg t_i$ and $w(t) \gg w(t_i)$, we expect that the influence of the initial condition  is negligible. In this particular limit,  \eqref{nice} should hold independently of the initial condition.

Introducing $\tu(t):=\int_{t'>t} \hu(t')$, \eqref{eq:EqUh1} gives the following equation for $\tu(t)$:
\begin{equation}
        \label{eq:EqUtPosBM}
        \partial_t \tu(t) - m^2 \tu(t) + \sigma \tu(t)^2 = - \int_{t'>t}\lambda(t'),
\end{equation}
where we used that $\tu(t)\to 0$ for $t\to + \infty$ (we recall that $\hat u(t)$ must vanish at
both $\pm \infty$).

\subsection{Example: Single-time position distribution}
To give a simple application of \eqref{eq:SolZPosStat}, we  compute the distribution of the position $u(t)$ at a single time. To do  this,  set $\lambda(t) = \lambda \delta(t-t_0)$ in \eqref{eq:MSRPathIntPos}. For the Brownian case, one obtains
\begin{eqnarray}
\nonumber
\lefteqn{\hu(t) =}&  &\\
\nonumber & & \frac{\lambda(1-4\lambda) 
\theta(t_0-t)}{\left\{\sinh\left[\frac{\sqrt{1-4\lambda}(t-t_0)}{2}\right]
-\sqrt{1-4\lambda}\cosh\left[\frac{\sqrt{1-4\lambda}(t-t_0)}{2}\right]\right\}^2}.
\end{eqnarray}
For the stationary case \eqref{eq:EqUtPosStat}, \(\hu(t)\) reads
\begin{eqnarray}
\nonumber
\hu(t) &=& \frac{\lambda\left(1-\lambda\right)^2 e^{-(t-t_0)(1-\lambda)} \theta(t_0-t)}{\left[e^{-(t-t_0)(1-\lambda)}-\lambda\right]^2}.
\end{eqnarray}
In both cases, the $\theta$ functions come from causality, since the driving $w(t)$ for $t>t_0$ cannot influence the measured position $u(t_0)$. Hence both $\tilde u(t)$ and $\hat u(t) = - \partial_t  \tilde u(t)$ must both be identically zero for $t>t_0$.

Let us assume a constant driving velocity, and write $w(t) = v (t-t_i) + w_i$. Then, for the one-sided Brownian
with $u(t_i)=0$ and $w_i \geq 0$ we have
\begin{eqnarray}
\nonumber
G(\lambda) &=& \overline{e^{\lambda u_{t_0}}} = e^{\int_{t_i}^{t_0} \rmd t' \hu(t') v t' } .
\end{eqnarray}
This leads to a complicated formula which simplifies in the limit $t_i \to -\infty$ at fixed $w(t_0)$,
\begin{equation}
G(\lambda) = \left(\frac{-2\lambda }{1-4\lambda-\sqrt{1-4\lambda}}\right)^{-v}e^{\frac{w(t_0)}{2}\left(1-\sqrt{1-4\lambda}\right)}.
\end{equation}
For the stationary case (restoring units), this is
\begin{eqnarray}
\nonumber
G(\lambda) &=& e^{m^2 \int_{t'}\hu(t') v t'\, dt'+\frac{\Delta(0)}{2 m^4}\lambda^2}\left(1- \frac{\sigma}{m^4} \lambda\right) \\
\nonumber
&=& \left(1-\frac{\sigma}{m^4} \lambda\right)^{\frac{m^2 v}{\sigma}+1}e^{\lambda v t_0+\frac{\Delta(0)}{2 m^4}\lambda^2}.
\end{eqnarray}
Inverting  gives a  valid distribution only for $| u_{t_0}- v t_0 | \ll \sigma/\Delta(0)$ which coincides with the
cut-off which should be used to regularize the stationary Brownian landscape (\ref{eq:DeltaStat}).

\section{Generalizations\label{sec:Generalizations}}
In light of the interesting results obtained for \eqref{eq:IntroABBM}, it is natural to ask whether our approach can be extended. In particular, one might want to replace the response function in \eqref{eq:IntroABBM} by a more general response kernel. For example, in order to model eddy currents which change the avalanche shape in real magnets \cite{ZapperiCastellanoColaioriDurin2005,Colaiori2008}, one may want to include second-order derivatives in time.

For this, it is useful to view the calculation in section \ref{sec:SolMSR} from another perspective. The equation \eqref{eq:EqUt} for $\tu$ is identical to the saddle-point equation obtained from the action \eqref{eq:MSRAction} in presence of the source $\lambda$ by taking a functional derivative with respect to $\du(t)$. The result \eqref{eq:SolZ} is then the value of $Z$ at the saddle point obtained by solving \eqref{eq:EqUt} for the given choice of $\lambda$. The other ``coordinate'' of the saddle point (which happens not to influence the value of $Z$ in this case, however) is the field $\du(t)$, fixed by the equation obtained by a functional derivative of \eqref{eq:MSRAction} with respect to $\tu(t)$,
\begin{equation}
\label{newu} 
\partial_t \du(t) + m^2\left[\du(t)-\dw(t)\right] - 2\sigma \du(t)\tu(t) = 0.
\end{equation}
This is the trajectory giving the dominant contribution to $Z$ for a given choice of $\lambda$. 
E.g. for $\lambda(t)=\lambda\delta(t-t_0)$, $\tu(t)$ is given by \eqref{eq:OneParticleUT}; for $w(t)=v t$ the solution of (\ref{newu}) converging to $v$ at infinity then reads
\begin{equation}
\nonumber
\du(t) = v\left(1+\frac{\lambda}{1-\lambda}e^{-|t-t_0|}\right).
\end{equation}
Note that it can also be obtained from the 2-time generating function (\ref{2timesZ}), e.g.~for $t>t_0$
as $\du(t)=\partial_{\lambda_2} \ln G(\lambda_1=\lambda,\lambda_2)|_{\lambda_2=0,t_2=t,t_1=t_0}$.
Indeed, since $S=\Gamma$ for monotonous motion, the solution of (\ref{newu}) identifies with (\ref{newu2}), i.e.~the saddle-point approximation is exact.
We thus see, as expected, that if we concentrate on small velocities ($\lambda\rightarrow -\infty$), the velocity on the dominant trajectory $\du(t)$ gets closer and closer to $0$ at $t_0$, but never becomes negative.

Now, the action $S$ generalizing \eqref{eq:MSRAction} with an arbitrary response kernel $R_{tt'}$ is
\begin{eqnarray}
\label{eq:MSRActionGen2}
\lefteqn{S[\du,\tu] =} & & \\
\nonumber
& & =\int_t \left\{ \tu(t)\left[\int_{t'} R^{-1}_{tt'} \du(t') -m^2\dw(t) \right] - \sigma \du(t) \tu(t)^2 \right\}.
\end{eqnarray}
The saddle-point equations read
\begin{eqnarray}
\nonumber
\int_{t'} R^{-1,T}_{tt'} \tu(t') - \sigma \tu(t)^2 - \lambda(t) &=& 0, \\
\label{eq:MSRSaddleGen}
\int_{t'} R^{ -1}_{tt'}\du(t') - 2\sigma \du(t)\tu(t) - m^2 \dw(t) &=& 0.
\end{eqnarray}
For a general (bare) response function $R$, the last term in the 
action (\ref{eq:MSRActionGen2}) is not exact, since we cannot assume monotonicity of each individual trajectory. However, as long as the saddle-point trajectory defined by \eqref{eq:MSRSaddleGen} for some choice of $\lambda$ is monotonous (i.e.~satisfies $\du(t)\geq 0$ for all $t$), it gives a well-defined approximation to the value of $Z$ for this particular $\lambda$. Investigating the quality of this approximation is an interesting avenue for further research.

\section{Summary and Outlook\label{sec:Conclusion}}
In this paper, we have considered the ABBM model with a monotonous, but non-stationary driving force. Using the Martin-Siggia-Rose formalism, we obtained the generating functional for the velocity from a field theory that can be solved exactly. This was illustrated on several paradigmatic examples (e.g.~a quench in the driving velocity). Using our formalism, we also succinctly recovered previous results on the stationary case.

An interesting direction for further research is trying to generalize these results to non-stationary dynamics of models which are not mean-field in nature, like $d$-dimensional elastic interfaces. Although some work has been done in that direction \cite{SchehrLeDoussal2005,KoltonSchehrLeDoussal2009,KoltonRossoGiamarchi2005,KoltonRossoAlbanoGiamarchi2006}, many questions remain open. Another complication arises when adding non-linear terms to the equation of motion \eqref{eq:IntroABBM} or \eqref{eq:EOMSpatial}. The effects of the KPZ term $[\nabla u(x)]^{2}$ have been discussed in \cite{CuleHwa1998,LeDoussalWiese2002a,ChenPapanikolaouSethnaZapperiDurin2011}. An anologous term but with a time- instead of a space derivative, i.e.\ a term $\dot u^2$, is related to dissipation of energy \footnote{If the equation of motion is $\partial _{t} u(t) = F(u(t))$, then $\int_{t} [\partial_{t}u(t)]^{2}  = \int_{t} F(u(t)) \partial_{t} u(t) = -\int _{t} \partial_{t} {\cal E}(u(t))$, where $F(u)=-\partial_{u} {\cal E}(u)$ . Thus this term, in the non-perturbed equation of motion, is related to dissipation of energy.} and yields a toy model with velocity-dependent friction. This is important as a step towards realistic earthquake models, where it is known that instead of a constant friction coefficient one has a complicated rate-and-state friction law \cite{Ruina1983,Dieterich1992,Scholz1998}. For the hysteresis loop in the ABBM model, it would  be  interesting to extend our results to the case of non-monotonous driving. Unfortunately, this is not  an easy task: We crucially used both the monotonicity of the particle velocity, $\du(t)\geq 0$, and the one of the driving, $\dw(t)\geq 0$ for simplifying the action and computing the path integral in section \ref{sec:SolMSR}. Without this assumption, neither the result \eqref{eq:SolZ} nor the non-renormalization property in section \ref{sec:NonRen} hold. Assuming the non-renormalization property, the mean velocity $\overline{\du(t)}$ would be equal to its value in the system without disorder at all times. This can be seen, e.g.~by taking $\partial_\lambda$ at $\lambda=0$ in formula (\ref{eq:DefZ}) and using (\ref{eq:SolZ}) and (\ref{eq:OneParticleUT}).
However, in numerical simulations one observes that this property breaks down as soon as the driving is non-monotonous, hence at least the term proportional to $\tu$ in the effective action is renormalized. We thus leave questions in this direction for future studies.

\acknowledgements We acknowledge Andrei Fedorenko for help in the derivation of the FRG equation (\ref{eq:FRGFlowDelta}).  This work was supported by ANR Grant No.~09-BLAN-0097-01/2, and by the CNRS through a doctoral fellowship for A.D.

\appendix
\section{Derivation of the non-stationary solution in discretized time\label{sec:AppDiscreteTime}}
The path integral derivation of \eqref{eq:SolZ} in section \ref{sec:SolABBM} is, to some extent, formal and neglects subtleties like convergence issues and boundary conditions. To complement it, we provide here a rigorous first-principle derivation of \eqref{eq:SolZ} by discretizing the time axis. For a small time step $\delta t$, we write \eqref{eq:SDEVel} as follows:
\begin{eqnarray}
\nonumber
\frac{\du_{j+1}-\du_j}{\delta t} &=& \frac{F\left(u_j + \delta t \du_{j+1}\right)-F(u_j)}{\delta t} \\
\nonumber
& & + m^2\left(\dw_{j+1} - \du_{j+1}\right) \\
\label{eq:DiscreteDu}
\Rightarrow \du_{j+1} &=& X(\du_{j+1}) + k m^2 \delta t \dw_{j+1} + k \du_j,
\qquad \end{eqnarray}
with $k^{-1}:=1+m^2\delta t$.\\
$X(\du_{j+1}) := k \left[F\left(u_j + \delta t \du_{j+1}\right)-F(u_j)\right]$ is, by the Markov property of Brownian motion, a new Brownian motion with $X(0)=0$ and variance $\overline{X(\du)X(\du')} = 2\sigma k^2 \delta t \min (\du,\du'$). Eq.~\eqref{eq:DiscreteDu} is an implicit equation for $\du_{j+1}$, which has, in general, several solutions $\du_{j+1}>0$. In fact, its solutions are the intersections of the Brownian motion $X(\du_{j+1})$ with the line $k m^2 \delta t \dw_{j+1} + k \du_j - \du_{j+1}$.
The true $\du_{j+1}$ describing the motion of the particle is the smallest of these solutions.  

Hence, the conditional probability distribution for $\du_{j+1}$ given $\du_j$ is  the first-passage distribution of Brownian motion, given by
\begin{equation}
P(\du_{j+1}|\du_j) = \frac{k m^2 \delta t \dw_{j+1} + k \du_j}{\sqrt{4\pi \sigma k^2 \delta t} \du_{j+1}^{\frac{3}{2}}} 
e^{-\frac{\left(\du_{j+1}- k m^2 \delta t \dw_{j+1} - k \du_j
\right)^2}{4\sigma k^2\delta t \du_{j+1}}}.
\end{equation}
The Laplace transform of this expression, which is the conditional expectation value for $e^{\tu\du_{j+1}}$, is given by
\begin{eqnarray}
\nonumber
 E(e^{\tu\du_{j+1}}|\du_j) &:=& \int_0^\infty e^{\tu\du_{j+1}} P(\du_{j+1}|\du_j) \rmd\du_{j+1}  \\
\label{eq:RecZ1}
&  =& e^{\frac{\dw_{j+1}m^2\delta t + \du_{j}}{2\sigma k \delta t}\left(1-\sqrt{1-4\tu \sigma k^2 \delta t }\right)}.\qquad 
\end{eqnarray}
This can be rewritten as
\begin{equation}
\label{eq:RecZ}
E(e^{\tu\du_{j+1}}|\du_j) = e^{m^2\tu'\dw_{j+1}\delta t} e^{\tu'\du_{j}},
\end{equation}
with $\tu'=\frac{1}{2\sigma k\delta t}\left(1-\sqrt{1-4\tu \sigma k^2\delta t}\right)$. 
Hence, iterating \eqref{eq:RecZ} one obtains
\begin{equation}
\label{eq:DiscreteZ}
\overline{e^{\sum_{j=1}^N \lambda_j \du_j \delta t}} = e^{m^2\sum_{j=1}^N \tu_j\dw_j \delta t}\times \overline{e^{\tu_1\du_0}},
\end{equation}
where $\tu_j$ is defined via the (backward) recursion
\begin{eqnarray}
\tu_{N+1} &=& 0 \\
\nonumber
\tu_j &=& \frac{1-\sqrt{1-4(\tu_{j+1}\delta t+\lambda_j \delta t^2) \sigma k^2}}{2\sigma k \delta t}, \quad 0 < j \leq N.
\end{eqnarray}
This is the exact solution for the discrete problem with $\delta t > 0$. In the continuum limit, we can take the leading order as $\delta t \rightarrow 0$. \eqref{eq:DiscreteZ} then reduces to the form \eqref{eq:SolZIC}. The recursion for $\tu$ becomes
\begin{equation}
\label{eq:RecursionUT}
\frac{\tu_j - \tu_{j+1}}{\delta t} = -m^2\tu_{j+1}  + \lambda_j + \sigma \tu_{j+1}^2 + \mathcal{O}(\delta t),
\end{equation}
which is  the discrete version of \eqref{eq:EqUt}. 

Let us now   show the connection with the MSR path integral discussed in section \ref{sec:SolMSR}.
We discretize the action \eqref{eq:MSRAction} with time step $\delta t$ using the It\^{o} prescription. Keeping $\du_j$ fixed, the path integral formula \eqref{eq:MSRPathInt} for the generating function \eqref{eq:DefZ} gives us the generating function for $\du_{j+1}$ as
%\begin{widetext}
\begin{align}
\label{eq:MSRDiscrete}
&E(e^{\lambda\du_{j+1}}|\du_j)= \int_{-\infty}^\infty \rmd\du_{j+1} \int_{-i\infty}^{i\infty} \frac{\rmd\tu_{j+1}}{2\pi} \\
&\qquad \qquad e^{-\tu_{j+1}\left[\frac{\du_{j+1}-\du_j}{\delta t} + m^2 \left(\du_j-\dw_{j}\right)\right]\delta t + \tu_{j+1}^2\sigma \du_{j} \delta t + \lambda \du_{j+1}}. \nn
\end{align}
%\end{widetext}
The integrals over $\tu_{j+1}$ and $\du_{j+1}$ can be performed explicitly, and yield (taking into account $\delta t>0$, $\sigma > 0$, $\dw \geq 0$, and $\du_j>0$)
\begin{equation}
E(e^{\lambda\du_{j+1}}|\du_j) = \exp\left[\left(\lambda-m^2 \delta t +\sigma \lambda^2\delta t\right)\du_j + \lambda m^2 \delta t \dw_j\right].
\end{equation}
To leading order for $\delta t \to 0$ and substituting $\lambda \to \tu$ this becomes identical to the generating function \eqref{eq:RecZ1}. Note that while the first-passage prescription used to obtain \eqref{eq:RecZ1} assumed $\du_{j+1}\geq 0$, in \eqref{eq:MSRDiscrete} we formally allow the velocity $\du_{j+1}$ to take any value between $-\infty$ and $\infty$. Surprisingly, this yields the same result to leading order in $\delta t$. It would be interesting to understand how a more rigorous MSR approach could be developed directly on the discrete version for finite $\delta t$ using first passage times. 

Analogously, one can derive a discretized path integral for the position variable $u$ for the one-sided Brownian potential discussed in section \ref{sec:PosFTBrownian}.

\begin{widetext}
\section{Functional determinants and 1-loop diagrams\label{sec:AppFDet}}
Here we  compute $\tr (R\Sigma)^n$ where $R$ is given in \eqref{eq:PropFree} and $\Sigma$ in \eqref{eq:DefSigma}. For simplicity we  set $\sigma=m=1$. Let us recall that in Ito discretization $\theta(0)=0$. First, note that
\begin{equation}
\nonumber
(R^T\Sigma^T)_{t_1,t_2} = \int_{t'} \hu(t') \sgn(t'-t_2) \left[\theta(t'-t_1)e^{-(t'-t_1)}-\theta(t_2-t_1)e^{-(t_2-t_1)}\right].
\end{equation}
Applying this to $\tr (R\Sigma)^n=\tr (R^T\Sigma^T)^n$, one gets
\begin{equation}
\nonumber
\tr (R\Sigma)^n = \int_{t'_1...t'_n} \hu(t'_1)...\hu(t'_n) \int_{t_1...t_n} \prod_{j=1}^n \sgn(t'_j-t_j)\left[\theta(t'_{j+1}-t_j)e^{-(t'_{j+1}-t_{j})}-\theta(t_{j+1}-t_{j})e^{-(t_{j+1}-t_{j})}\right].
\end{equation}
The convention is that $t_{n+1}=t_1$ and $t'_{n+1}=t'_1$.
Now, we conjecture that for any $t'_1...t'_n$, 
\begin{equation}
\label{eq:TraceConjecture}
 \int_{t_1...t_n} \prod_{j=1}^n \sgn(t'_j-t_j)\left[\theta(t'_{j+1}-t_j)e^{-(t'_{j+1}-t_{j})}-\theta(t_{j+1}-t_{j})e^{-(t_{j+1}-t_{j})}\right]
=(-1)^{n+1} .
\end{equation}
We were unable to find an analytic proof, but verified this conjucture for $n\leq 5$. Assuming it for any $n$, one obtains as claimed
\begin{equation}
\nonumber
\tr (R\Sigma)^n = -\left[-\int_t \hu(t)\right]^n.
\end{equation}
For the one-sided Brownian correlator (\ref{eq:DeltaBM}) we find the self-energy analogous to
(\ref{eq:DefSigma}) as
\begin{eqnarray} 
\Sigma_{t_1,t_2} = - 2 \delta(t_1-t_2) \int_{t'} \hat u(t') \theta(t'-t_2) - 2 \hat u(t_2) \theta(t_1-t_2). 
\end{eqnarray} 
This implies
\begin{equation}
\nonumber
(R^T\Sigma^T)_{t_1,t_2} = -2 \int_{t'} \hu(t')  \left[\theta(t_2-t') \theta(t'-t_1)e^{-(t'-t_1)} + \theta(t'-t_2)\theta(t_2-t_1)e^{-(t_2-t_1)}\right].
\end{equation}
One then finds $\tr (R\Sigma)^n =0$ for $n \geq 1$ hence a unit functional determinant as claimed in the text.

This can be generalized to the $d$-dimensional interface. We need to compute  the functional determinant
$\det( 1 + R \Sigma )$ with
\begin{eqnarray}
&& R^{-1}_{x_1 t_1,x_2 t_2} = \delta_{t_1 t_2} ( \partial_{t_2} \delta_{x_1 x_2} + g_{x_1 x_2}  )  \\
&& \Sigma_{x_1 t_1,x_2 t_2}  = \delta_{x_1 x_2} [  \sigma \delta_{t_1 t_2} \int_{t'} \hu_{x t'} \text{sgn}(t_1-t') +\sigma \hu_{x t_2}  \text{sgn}(t_2-t_1) ]. 
\end{eqnarray}
We conjecture that this yields
\begin{eqnarray}
\ln \det(1 + R \Sigma) = \tr \ln\left( \delta_{x x'}  - \sigma g_{xx'} \int_t \hat u_{x't} \right) = 
\tr \ln\left[ \delta_{x x'}  - \sigma g_{xx'} \int_{y} g_{x' y} \int_t \lambda_{yt} \right].
\end{eqnarray} 
For the last equality, we used 
$\int_t \hat u_{x t} = g_{xx'}  \int_t \lambda_{x't}$. For a uniform source one recovers the expression in the text of section \ref{sec:PosFTStat}. 

\end{widetext}
\section{1-loop functional RG at finite velocity}
\label{frg}

In  \cite{ChauveGiamarchiLeDoussal2000}, the 1-loop functional RG equations for a $d$-dimensional elastic interface at non-zero driving velocity $v>0$ were derived in the Wilson RG scheme. These equations have resisted analytical (or numerical) solution since then. 
Here, instead of using Wilson RG with a hard cutoff in momentum space, we regularize our model by a parabolic well with curvature $m^2$. We point out that the stationary ABBM disorder correlator \eqref{defDE}, \eqref{eq:DeltaStat} yields a simple solution of the corresponding functional RG equations. This also provides an independent check of the non-renormalization property for ABBM disorder discussed in section \ref{sec:NonRen} using a  different method.

For a $d$-dimensional interface driven by a parabolic well of curvature $m^2$ centered at $w=v t$, one can derive the functional RG flow equation by computing $- m \partial_m \Gamma$ and reexpressing it as a function of $\Gamma$. This is done order by order in $\Delta$, which in this Appendix denotes the {\it renormalized} second cumulant of the disorder (the local part of the term $\hat u \hat u$ in $\Gamma$).
The resulting functional RG flow of $\Delta$ at finite driving velocity $v$ is \cite{FedorenkoLeDoussalWiese2008}
\begin{widetext} 
\begin{eqnarray}
\nonumber
-m\partial_m \tD(u)&=&(\epsilon-2\zeta) \tD(u) + \zeta u \tD'(u) \nn \\ \nn &&+ \int_0^\infty \rmd s_1 \int_0^\infty \rmd s_2 \frac{e^{-(s_1+s_2)}}{s_1+s_2}\left\{\tD''(u)\left[\tD(\tilde v(s_2-s_1))-\tD(u+\tilde v(s_2-s_1))\right]\right. \\
\label{eq:FRGFlowDelta} & & \left.-\tD'(u+\tilde v s_1)\tD'(u-\tilde vs_2)+\tD'(\tilde v(s_1+s_2))\left[\tD'(u-\tilde vs_1)-\tD'(u+\tilde vs_2)\right]\right\}.
\end{eqnarray}
\end{widetext}
Here $\epsilon=4-d$, the rescaled correlator is defined via $\Delta(u)  = A_d m^{\epsilon - 2 \zeta} \tD(u m^\zeta)$
with $A_d^{-1} = \epsilon \int \frac{d^d k}{(2 \pi)^d} (1+k^2)^{-2}$, and  $\tilde  v = \eta_m v/m^{2-\zeta}$ flows as 
\begin{eqnarray}
- m \partial_m \ln \tilde v = z-\zeta = 2 - \zeta - \int_{s>0} e^{-s}  \tD''(s \tilde v).
\end{eqnarray}
The flow of $\tilde v$ arises because the friction is corrected by disorder. In general, this leads to a non-trivial dynamical exponent $z$ defined by the relation above. For $v \to 0$ one recovers the flow at the depinning threshold obtained in \cite{LeDoussalWieseChauve2002}. These equations are sufficient \footnote{Generally one should write the flow of all terms in $\Gamma$, e.g.~flow of the inverse response function $R^{-1}$ to $O(\tD)$, but to lowest order in $\epsilon$ it is sufficient to consider only the friction $\eta_m$. Similarly the higher cumulants of the disorder, and the non-local part of the second cumulant, are of higher order in $\epsilon$.} for an expansion in
$\epsilon$ with $\tD = O(\epsilon)$.

Plugging in the correlator for ABBM-type disorder, $\tD(u) = \tD(0) - \tilde \sigma |u|$, 
and $\zeta=\epsilon$ into \eqref{eq:FRGFlowDelta}, one finds 
\begin{eqnarray}
 -m\partial_m \Delta(u) &=& - \tilde \sigma^2 \\
 -m\partial_m \tilde v &=& z-\zeta = 2 - \epsilon.
\end{eqnarray}
We see that the dynamical exponent $z$ for ABBM-type disorder takes the value $z=2$ in any dimension $d$. The ABBM form of the disorder is preserved with $- m \partial_m \tilde \sigma=0$ and only $\tD(0)$ 
flowing as $- m \partial_m \tD(0)= - \tilde \sigma^2$. This is consistent (for $d=0$) with equation
(\ref{eq:SolZPosStat}). In addition, as discussed in section \ref{sec:FTPos} and appendix \ref{sec:AppFDet}, 2- and higher-loop corrections vanish in any $d$ for monotonous motion in ABBM-type disorder. More precisely, $\Gamma - S$ is the logarithm of a functional determinant 
computed in section \ref{sec:FTPos}. This shows that for ABBM-type disorder, \eqref{eq:FRGFlowDelta} is exact to all orders in $\epsilon=4-d$. 

We note that for ABBM disorder the correlator remains non-analytic for any $v$
\footnote{This may appear to be in contradiction to the discussion in section VI B of \cite{LeDoussalWiese2009}. Note however, that the $\Delta$ appearing in Eq.~(226) of \cite{LeDoussalWiese2009} is defined as a 
two-point correlation function of $u$. The $\Delta$ we compute here is the $\hu^2$ term in $\Gamma$. It remains non-analytic at finite $v$, but to go to the correlation function one needs to convolve it with two propagators. This smoothens the linear cusp to the sub-cusp discussed in \cite{LeDoussalWiese2009}.}. This is, presumably, a 
peculiarity of ABBM disorder. For short-ranged disorder  this may only hold until some scale, the non-analyticity being rounded at larger scales (small $m$). However further studies are needed to  clarify the validity of this hypothesis.

\bibliography{msrabbm}

%\tableofcontents

\end{document}